\documentclass[notitlepage,onecolumn,10pt]{revtex4-1}
\usepackage[english]{babel}
\usepackage[latin1]{inputenc}
\usepackage{amsmath}
\usepackage{amssymb}
\usepackage{bbm}
\usepackage{graphics}
\begin{document}
\title{Effective Chern-Simons actions of particles coupled to 3D gravity}
\author{Tomasz Trze\'{s}niewski
}
\email{tbwbt@ift.uni.wroc.pl}
\affiliation{Institute for Theoretical Physics, University of Wroc\l{}aw, pl.\ 
M.\ Borna 9, 50-204 Wroc\l{}aw, Poland}
\date{\today}

\begin{abstract}
Point particles in 3D gravity are known to behave as topological defects, while gravitational field can be expressed as the Chern-Simons theory of the appropriate local isometry group of spacetime. In the case of the Poincar\'{e} group, integrating out the gravitational degrees of freedom it is possible to obtain the effective action for particle dynamics. We review the known results, both for single and multiple particles, and attempt to extend this approach to the (anti-)de Sitter group, using the factorizations of isometry groups into the double product of the Lorentz group and ${\rm AN}(2)$ group. On the other hand, for the de Sitter group one can also perform a contraction to the semidirect product of ${\rm AN}(2)$ and the translation group. The corresponding effective action curiously describes a Carrollian particle with the ${\rm AN}(2)$ momentum space. We derive this contraction in a more rigorous manner and further explore its properties, including a generalization to the multiparticle case.
\end{abstract}

\maketitle

\section{Introduction} \label{sec:1.0}
Gravity in 2+1-dimensional spacetime offers an attractive combination of the conceptual foundations descending from ordinary general relativity and the apparent physical simplicity. It cannot accommodate local degrees of freedom: there are no forces acting at a distance and no gravitational waves. Consequently, spacetime is locally isometric to flat Minkowski space or constantly curved (anti-)de Sitter space, depending on the cosmological constant. The theory can be equipped with the topological degrees of freedom, by considering a spatial topology with handles (not discussed in this paper) or including point particles, which themselves turn out to be topological defects with the geometry of a cone \cite{Staruszkiewicz:1963ge,Deser:1984tc,Deser:1984tr}. Namely, an embedding of a neighbourhood of such a defect is constructed by taking Minkowski or, respectively, (anti-)de Sitter space and removing a wedge whose edge is the particle's worldline, while the defect's deficit angle is determined by the particle's relativistic mass. To recover a conical spacetime the faces of the wedge have to be identified by the holonomy of a loop encircling the defect, which is a Lorentz transformation conjugate to a rotation by the deficit angle. This Lorentz group element can actually be interpreted \cite{Matschull:1997qy,Matschull:2001ty} as the particle's physical momentum. Similarly, the effect of non-zero spin of the particle (as well as the orbital angular momentum) is a dislocation along the wedge, associated with a time translation in the holonomy, which generalizes the defect's geometry to that of a helical cone \cite{Deser:1984tc}. Spinning particles may therefore lead to the occurrence of closed timelike curves but this problem will hopefully be resolved in the full quantum theory. 

Due to the properties of possible local isometry groups of spacetime, gravity in three dimensions can also be formulated \cite{Achucarro:1986as,Witten:1988dm} as the Chern-Simons gauge theory (although there are certain subtleties to be taken into account). In principle, if conical defects are coupled to such a theory, redundant gravitational degrees of freedom can be integrated out to obtain the effective dynamics of particles. In the case of the Poincar\'{e} gauge group this can be accomplished in the language of the symplectic form \cite{Meusburger:2005by} or the corresponding action \cite{Arzano:2014by}. However, to our knowledge, so far it has not been done for the (anti-)de Sitter group. One of the results of the present paper is an explicit calculation of the effective particle actions in the latter cases, although they still require a very complicated integration. To this end, for each of the gauge groups we use here the same local factorization \cite{Meusburger:2008qr} into the product of the (three-dimensional) Lorentz group and the so-called ${\rm AN}_{\bf n}(2)$ group. 

What is particularly interesting about such a factorization is that ${\rm AN}(2)$ (with the timelike deformation vector ${\bf n}$) is the three-dimensional counterpart of the ${\rm AN}(3)$ group, which plays the role \cite{Majid:1994by} of covariant momentum space under the action of the $\kappa$-Poincar\'{e} (Hopf) algebra \cite{Lukierski:1991qa,Lukierski:1992ny} -- the best studied quantum deformation of the Poincar\'{e} algebra. Deformations of relativistic symmetries are conjectured to arise in certain approximations to the quantum theory of gravity and have especially been considered in the framework known as doubly (or deformed) special relativity, later recast under the name of relative locality, which is motivated by the phenomenological speculations \cite{Amelino:2011py,Gubitosi:2013re}. For these reasons a natural question is whether the (three-dimensional) $\kappa$-Poincar\'{e} algebra, or its (anti-)de Sitter analogue \cite{Ballesteros:1994qh,Ballesteros:2004ns}, can appear as a description of symmetries in the context of quantum gravity in three dimensions, as suggested in \cite{Amelino:2004qy}. A conclusion of the rigorous analysis \cite{Meusburger:2009gy} was that there is no Chern-Simons action associated with the $\kappa$-deformed relativistic symmetries and simultaneously equipped with the scalar product corresponding to three-dimensional gravity. However, a loophole in this argument has recently been found \cite{Rosati:2017dt}, which apparently allows the desired symmetries to arise. It has also been shown (at least in the Euclidean domain) \cite{Cianfrani:2016ss} that one obtains the $\kappa$-Poincar\'{e} algebra by loop-quantizing the algebra of gravitational constraints with positive cosmological constant and subsequently performing a contraction of such a quantum algebra. 

On the other hand, with a similar motivation, in \cite{Kowalski:2014dy} we introduced a contraction of the Chern-Simons theory with the de Sitter gauge group that is in a certain sense opposite to the standard contraction leading to the theory with the Poincar\'{e} group. It gives us the effective action with the ${\rm AN}(2)$ particle momentum space. However, the action actually describes the peculiar $\kappa$-deformed version of a Carroll (or ultralocal) particle instead of a $\kappa$-deformed relativistic particle. In the present paper we explain the above contraction (which we call here the reciprocal contraction) in more detail, implement it more rigorously and add a discussion of the obtained gauge algebra and the particle's holonomy, as well as we consider the generalization to multiparticle case. Moreover, we find that the analogous contractions for the anti-de Sitter group and the double-product factorization of the Poincar\'{e} group lead to diverging Chern-Simons actions. 

This paper has the following contents. In the next Section~\ref{sec:2.0} we briefly introduce the Chern-Simons action describing the dynamics of three-dimensional gravity with a single particle and begin the procedure that leads to the effective particle action. The calculations are continued for the de Sitter gauge group in Subsection~\ref{sec:2.1}, where we also recover the final action in the Poincar\'{e} case. The Subsection~\ref{sec:2.2} is devoted to a derivation of the action for the reciprocal contraction of the de Sitter group, while in Subsection~\ref{sec:2.3} we discuss the cases of anti-de Sitter as well as the factorized Poincar\'{e}. Subsequently, in Section~\ref{sec:3.0} we generalize our considerations to multiple particles. Subsections \ref{sec:3.1} and \ref{sec:3.2} contain an analysis of the Poincar\'{e} and $\kappa$-Carroll cases, respectively. The essence of results concerning the Poincar\'{e} gauge group is well known but here we try to present them in a very transparent manner and include massless particles. After a brief summary and some outlook for the future research, in the Appendix we collect the mathematical knowledge concerning all the gauge groups, including the explicit form of some formulae that were not written down in earlier papers.

\section{Single particle} \label{sec:2.0}
As we already mentioned in the Introduction, the special nature of gravity in 2+1 spacetime dimensions allows to formulate it as the Chern-Simons gauge theory. The appropriate gauge group for this purpose is the group of local isometries of spacetime, which is the (double cover of the) three-dimensional Poincar\'{e}, de Sitter or anti-de Sitter Lie group, respectively for zero, positive or negative cosmological constant $\Lambda$. The corresponding Lie algebra can always be written in terms of the generators $J_\alpha$, $P_\alpha$, $\alpha = 0,1,2$ with the brackets
\begin{align}\label{eq:20.01}
[J_\alpha,J_\beta] = \epsilon_{\alpha\beta\gamma} J^\gamma\,, \qquad 
[J_\alpha,P_\beta] = \epsilon_{\alpha\beta\gamma} P^\gamma\,, \qquad 
[P_\alpha,P_\beta] = -\Lambda \epsilon_{\alpha\beta\gamma} J^\gamma\,,
\end{align}
where we set the conventions for the metric $\eta = {\rm diag}(1,-1,-1)$ and Levi-Civita symbol $\epsilon_{012} = 1$ (for more details see the Appendix). There exists a two-dimensional space of possible scalar products on the above algebra but to obtain the correct gravitational action we have to use the one of the form \cite{Witten:1988dm}
\begin{align}\label{eq:20.02}
\left<J_\alpha, P_\beta\right> = \eta_{\alpha\beta}\,, \qquad 
\left<J_\alpha, J_\beta\right> = \left<P_\alpha, P_\beta\right> = 0\,.
\end{align}
The gauge field of this theory is the Cartan connection, which is an algebra-valued one-form
\begin{align}\label{eq:20.03}
A = \omega^\alpha J_\alpha + e^\alpha P_\alpha\,,
\end{align}
constructed from the spin connection $\omega^\alpha = \omega^\alpha_\mu dx^\mu$ and dreibein $e^\alpha = e^\alpha_\mu dx^\mu$ (assumed to be invertible). Its curvature is given by $F = dA + A \wedge A = R + T + C$, where
\begin{align}\label{eq:20.03a}
R = \left(d\omega^\alpha + \frac{1}{2}\, \epsilon^\alpha_{\ \beta\gamma} \omega^\beta \wedge \omega^\gamma\right) J_\alpha\,, \qquad T = \left(de^\alpha + \epsilon^\alpha_{\ \beta\gamma} \omega^\beta \wedge e^\gamma\right) P_\alpha\,, \qquad C = \frac{\Lambda}{2}\, \epsilon^\alpha_{\ \beta\gamma} e^\beta \wedge e^\gamma J_\alpha
\end{align}
are respectively the Riemann curvature, torsion and cosmological constant term. 

In order to explore the dynamics of point particles coupled to the gravitational field let us assume that spacetime has the product structure $\mathbbm{R} \times {\cal S}$, where ${\cal S}$ is a spatial submanifold of genus $0$. We accordingly decompose the connection into $A = A_t dt + A_{\cal S}$, with a one-form $A_{\cal S}$ defined on ${\cal S}$ and the spatial curvature $F_{\cal S} = dA_{\cal S} + A_{\cal S} \wedge A_{\cal S}$. A particle can be introduced as a (spinning) conical singularity, appearing as a puncture on ${\cal S}$. When ${\cal S}$ has the open topology (i.e.\! for $\Lambda \leq 0$), we restrict to the connections $A_{\cal S}$ that satisfy such fall-off conditions at spatial infinity that the geometry of spacetime is asymptotically conical. Then in the single particle case the total Chern-Simons action of the system has the form \cite{Witten:1989ql,Sousa:1990os,Meusburger:2009gy}
\begin{align}\label{eq:20.04}
S = \int dt L\,, \quad L &= \frac{k}{4\pi} \int_{\cal S} \left<\dot A_{\cal S} \wedge A_{\cal S}\right> - 
\left<c_0 h^{-1} \dot h\right> \nonumber\\ 
&+ \int_{\cal S} \left<A_t \left(\frac{k}{2\pi} F_{\cal S} - 
h c_0 h^{-1} \delta^2(\vec{x} - \vec{x}_*)\, dx^1 \wedge dx^2\right)\right>\,,
\end{align}
where $k = 1/(4G)$ is the coupling constant and $x^1,x^2$ denote coordinates on ${\cal S}$ with the origin at the particle's position $\vec{x}_*$. Besides, the gauge algebra element $c_0 = c_J + c_P \equiv m\, J_0 + s\, P_0$ encodes mass $m > 0$ and spin $s$ of the particle, while $h$ is a group element that through the conjugation $h c_0 h^{-1} = p^\alpha J_\alpha + j^\alpha P_\alpha$ determines the particle's momentum ${\bf p} = p^\alpha J_\alpha$ and generalized angular momentum ${\bf j} = j^\alpha P_\alpha$. In particular, in the case $\Lambda = 0$ we have the standard relation $j^\alpha = \epsilon^\alpha_{\ \beta\gamma} x^\beta p^\gamma + s\, \hat p^\alpha$ (with $\hat p^\alpha \equiv p^\alpha/m$ and $x^0 \equiv t$). The first term of the Lagrangian in (\ref{eq:20.04}) describes the gravitational field, the second one a free particle and the last one their mutual interaction. Treating $A_t$ as a Lagrange multiplier we can interpret the latter term as a constraint on the spatial curvature:
\begin{align}\label{eq:20.05}
\frac{k}{2\pi} F_{\cal S} = h c_0 h^{-1} \delta^2(\vec{x} - \vec{x}_*)\, 
dx^1 \wedge dx^2\,.
\end{align}
From $F_{\cal S} = R_{\cal S} + T_{\cal S} + C_{\cal S}$ it then follows that the (spatial) Riemann curvature and torsion are given by
\begin{align}\label{eq:20.06}
R_{\cal S} = -C_{\cal S} + \frac{2\pi}{k} {\bf p}\, \delta^2(\vec{x} - \vec{x}_*)\, dx^1 \wedge dx^2\,, \qquad 
T_{\cal S} = \frac{2\pi}{k} {\bf j}\, \delta^2(\vec{x} - \vec{x}_*)\, dx^1 \wedge dx^2\,.
\end{align}
They both vanish (on the background of constant curvature $R_{\cal S} = -C_{\cal S}$) everywhere except the puncture, where the momentum ${\bf p}$ is a source of curvature and the generalized angular momentum ${\bf j}$ a source of torsion. 

The constraint (\ref{eq:20.05}) allows us to gauge away the gravitational connection by expressing it in terms of the particle degrees of freedom, which leads to the effective particle action. To this end we will employ the approach introduced in \cite{Alekseev:1995se,Meusburger:2005by,Meusburger:2006pe}, where it was used to derive the corresponding symplectic form. The first step is to divide the spatial slice ${\cal S}$ into a region containing the particle ${\cal D}$, topologically equivalent to a punctured disc, with polar coordinates $\rho \in (0,1]$, $\phi \in [0,2\pi]$, and the remaining empty region ${\cal E}$ (where $\rho \geq 1$). They are separated by the circular boundary $\Gamma$ (at $\rho = 1$). On the empty region ${\cal E}$ the spatial connection is flat (even for $\Lambda \neq 0$) and has the general form
\begin{align}\label{eq:20.07}
A_{\cal S}^{({\cal E})} = \gamma d\gamma^{-1}\,,
\end{align}
where $\gamma$ is a certain gauge group element. Solving the constraint (\ref{eq:20.05}) on the punctured disc ${\cal D}$ we similarly find that the connection is given by
\begin{align}\label{eq:20.08}
A_{\cal S}^{({\cal D})} = \bar\gamma\, \frac{c_0}{k} d\phi\, \bar\gamma^{-1} + 
\bar\gamma d\bar\gamma^{-1}\,, \quad \bar\gamma(\rho = 0) = h\,,
\end{align}
where a group element $\bar\gamma$ is associated with the particle's motion and for an infinitesimally small ${\cal D}$ it reduces to $h$. Furthermore, requiring continuity of $A_{\cal S}$ across the boundary $\Gamma$, i.e.\! $A_{\cal S}^{({\cal D})}|_\Gamma = A_{\cal S}^{({\cal E})}|_\Gamma$, is equivalent to the sewing condition
\begin{align}\label{eq:20.09}
\gamma^{-1}|_\Gamma = \alpha\, e^{\frac{1}{k} c_0 \phi}
\bar\gamma^{-1}|_\Gamma\,,
\end{align}
with an arbitrary constant group element $\alpha = \alpha(t)$, $d\alpha = 0$. The jump of the value of $\gamma$ (while $\bar\gamma$ is continuous) at the point $\phi = 2\pi$, which coincides with $\phi = 0$, is an effect of the conical singularity at $\rho = 0$, characterized by the nontrivial holonomy of $A_{\cal S}^{({\cal E})}$ (see Subsection~\ref{sec:2.1}). 

As we describe it in the Appendix, every gravitational gauge group can be locally factorized into the product of groups ${\rm SL}(2,\mathbbm{R})$ and ${\rm AN}_{\bf n}(2)$, with the deformation vector ${\bf n} \in \mathbbm{R}^{2,1} \backslash \{0\}$, ${\bf n}^2 = \Lambda$ \cite{Meusburger:2008qr}. ${\bf n}$ is timelike in the de Sitter, lightlike in the Poincar\'{e} and spacelike in the anti-de Sitter case. The factorized group has the double product structure ${\rm SL}(2,\mathbbm{R}) \vartriangleright\!\!\vartriangleleft {\rm AN}_{\bf n}(2)$, where both subgroups are acting on each other in a complicated manner. However, in terms of the $\mathfrak{an}_{\bf n}(2)$ generators
\begin{align}\label{eq:20.10}
S_\alpha = P_\alpha + \epsilon_{\alpha\beta\gamma} n^\beta J^\gamma\,,
\end{align}
the scalar product (\ref{eq:20.02}) on the corresponding algebra simply becomes
\begin{align}\label{eq:20.11}
\left<J_\alpha, S_\beta\right> = \eta_{\alpha\beta}\,, \qquad 
\left<J_\alpha, J_\beta\right> = \left<S_\alpha, S_\beta\right> = 0\,.
\end{align} 
We choose here the order of factorization in which gauge group elements $g$ are expressed as
\begin{align}\label{eq:20.12}
g = \mathfrak{u}\, \mathfrak{s} = (u_3 \mathbbm{1} + u^\alpha J_\alpha) 
(s_3 \mathbbm{1} + s^\beta S_\beta)\,,
\end{align}
under the factorization condition $s_3 + \tfrac{1}{2} {\bf n} \cdot {\bf s} > 0$. The choice of the reverse ordering (\ref{eq:A1.11}) would lead to deriving the effective action that differs by the appropriate group conjugations. 

Substituting the decomposed connection (\ref{eq:20.07}-\ref{eq:20.08}) into (\ref{eq:20.04}) and factorizing both $\gamma$ and $\bar\gamma$, we can rewrite the Lagrangian in the boundary form
\begin{align}\label{eq:20.13}
L = \frac{k}{2\pi} \int_\Gamma \left<d\mathfrak{s}\, \mathfrak{s}^{-1} 
\mathfrak{u}^{-1} \dot{\mathfrak{u}} - d\bar{\mathfrak{s}}\, 
\bar{\mathfrak{s}}^{-1} \bar{\mathfrak{u}}^{-1} \dot{\bar{\mathfrak{u}}} + 
\frac{c_0}{k} d\phi \left(\bar{\mathfrak{s}}^{-1} \bar{\mathfrak{u}}^{-1} 
\dot{\bar{\mathfrak{u}}}\, \bar{\mathfrak{s}} + \bar{\mathfrak{s}}^{-1} 
\dot{\bar{\mathfrak{s}}}\right)\right>\,,
\end{align}
where the contribution from the disc ${\cal D}$ had to be included with the opposite orientation of $\Gamma$. Subsequently, with the help of the sewing condition (\ref{eq:20.09}) we can eliminate $d\mathfrak{s}\, \mathfrak{s}^{-1}$ from (\ref{eq:20.13}) to obtain
\begin{align}\label{eq:20.14}
L = \frac{k}{2\pi} \int_\Gamma \left<\partial_0\left(\bar{\mathfrak{u}}^{-1} 
\mathfrak{u}\right) \mathfrak{u}^{-1} \bar{\mathfrak{u}} 
\left(d\bar{\mathfrak{s}}\, \bar{\mathfrak{s}}^{-1} - 
\bar{\mathfrak{s}}\, \frac{c_0}{k} d\phi\, \bar{\mathfrak{s}}^{-1}\right) + \frac{c_0}{k} d\phi\, \bar{\mathfrak{s}}^{-1} 
\dot{\bar{\mathfrak{s}}}\right>\,.
\end{align}
We will proceed further with this expression in the specific cases.

\subsection{De Sitter and Poincar\'{e} gauge groups} \label{sec:2.1}
Let us first consider the Lagrangian (\ref{eq:20.14}) for $\Lambda > 0$, with the de Sitter gauge group ${\rm SL}(2,\mathbbm{C})$, and take ${\bf n} = (\sqrt{\Lambda},0,0)$ as the deformation vector. Strictly speaking, a single particle solution for positive $\Lambda$ can not exist since in this case the spherical topology of ${\cal S}$ requires the presence of a complementary conical defect \cite{Deser:1984tr}. However, here we introduce it as a step towards either the multiparticle case or contractions of the theory with the ${\rm SL}(2,\mathbbm{C})$ group. We have $S_0 = P_0$, $c_S \equiv c_P$ and therefore the group element characterizing a particle in (\ref{eq:20.09}) has the exceptionally simple factorization
\begin{align}\label{eq:21.01}
e^{\frac{1}{k} c_0 \phi} = e^{\frac{1}{k} c_J \phi} e^{\frac{1}{k} c_S \phi} = e^{\frac{1}{k} c_S \phi} e^{\frac{1}{k} c_J \phi} = (\sigma_3 \mathbbm{1} + \sigma_0 S_0) (\mu_3 \mathbbm{1} + \mu_0 J_0)\,,
\end{align}
where $\sigma_3 \pm \tfrac{1}{2} \sqrt{\Lambda}\, \sigma_0 > 0$ is automatically satisfied, since
\begin{align}\label{eq:21.02}
\mu_3 &= \cos\frac{m \phi}{2k}\,, & \mu_0 &= 2 \sin\frac{m \phi}{2k}\,, \nonumber\\
\sigma_3 &= \cosh\frac{\sqrt{\Lambda}\, s \phi}{2k}\,, & \sigma_0 &= \frac{2}{\sqrt{\Lambda}}\, \sinh\frac{\sqrt{\Lambda}\, s \phi}{2k}\,.
\end{align}
It is suitable to choose the reverse-ordered factorization (\ref{eq:A1.11}) of the constant group element
\begin{align}\label{eq:21.03}
\alpha = \mathfrak{r}\, \mathfrak{v} = (r_3 \mathbbm{1} + r^\alpha S_\alpha) 
(v_3 \mathbbm{1} + v^\beta J_\beta)\,,
\end{align}
with $r_3 - \tfrac{1}{2} \sqrt{\Lambda}\, r_0 > 0$. Applying (\ref{eq:21.01}) and (\ref{eq:21.03}) to the sewing condition (\ref{eq:20.09}) and using the formulae (\ref{eq:A1.12}-\ref{eq:A1.13}) from the Appendix, after lengthy calculations we find that the explicit expression for $\mathfrak{u}^{-1}$ is given by
\begin{align}\label{eq:21.04}
\mathfrak{u}^{-1} = (V_3 \mathbbm{1} + V^\alpha J_\alpha)\, \bar{\mathfrak{u}}^{-1}\,, \quad
V_3 &= \frac{1}{N_V} \left(U_3 \bar s_3 + \tfrac{1}{2} \sqrt{\Lambda} (U_3 \bar s_0 + \epsilon_{0\alpha\beta} U^\alpha \bar s^\beta)\right)\,, \nonumber\\
V^\alpha &= \frac{1}{N_V} \left((\bar s_3 - \tfrac{1}{2} \sqrt{\Lambda}\, \bar s_0) U^\alpha + \sqrt{\Lambda}\, U^\beta \bar s_\beta \eta^{\alpha 0}\right)\,, \nonumber\\
N_V^2 &\equiv (\bar s_3 + \tfrac{1}{2} \sqrt{\Lambda}\, \bar s_0)^2 + \tfrac{1}{4} \Lambda\, \bar s^\alpha \bar s_\alpha (U_1^2 + U_2^2) \nonumber\\
&+ \sqrt{\Lambda} (\bar s_3 + \tfrac{1}{2} \sqrt{\Lambda}\, \bar s_0) \left(\tfrac{1}{2} (U^\alpha \bar s_\alpha U_0 - U^\alpha U_\alpha \bar s_0) + U_3 \epsilon_{0\alpha\beta} U^\alpha \bar s^\beta\right)\,,
\end{align}
which is valid when $N_V^2 > 0$ and where
\begin{align}\label{eq:21.05}
U_3 &= \frac{1}{N_U} (\mu_3 v_3 - \tfrac{1}{4} \mu_0 v_0) (\sigma_3 - \tfrac{1}{2} \sqrt{\Lambda}\, \sigma_0)\,, \nonumber\\
U^\alpha &= \frac{1}{N_U} \left((\sigma_3 + \tfrac{1}{2} \sqrt{\Lambda}\, \sigma_0) (\mu_3 v^\alpha + \mu_0 v_3 \eta^{\alpha 0} + \tfrac{1}{2} \epsilon^\alpha_{\ \beta 0} \mu_0 v^\beta) - (\mu_3 v_0 + \mu_0 v_3) \sqrt{\Lambda}\, \sigma_0 \eta^{\alpha 0}\right)\,, \nonumber\\
N_U &\equiv \sqrt{(\sigma_3 - \tfrac{1}{2} \sqrt{\Lambda}\, \sigma_0)^2 + \tfrac{1}{2} \sqrt{\Lambda}\, \sigma_3 \sigma_0 (v_1^2 + v_2^2)}\,,
\end{align}
which has to satisfy $N_U^2 > 0$. Substituting (\ref{eq:21.04}) into (\ref{eq:20.14}) we obtain the effective Lagrangian
\begin{align}\label{eq:21.06}
L = \frac{k}{2\pi} \int_\Gamma \left<f(V) \left(d\bar{\mathfrak{s}}\, \bar{\mathfrak{s}}^{-1} - \bar{\mathfrak{s}}\, \frac{c_0}{k} d\phi\, \bar{\mathfrak{s}}^{-1}\right) + \frac{c_0}{k} d\phi\, \bar{\mathfrak{s}}^{-1} 
\dot{\bar{\mathfrak{s}}}\right>\,, \nonumber\\
f(V) \equiv \left(\dot V_3 V^\alpha - V_3 \dot V^\alpha + \tfrac{1}{2} \epsilon^\alpha_{\ \beta\gamma} V^\beta \dot V^\gamma\right) J_\alpha\,.
\end{align}
As the last step we should perform the integration in (\ref{eq:21.06}) over the boundary coordinate $\phi \in [0,2\pi]$. Unfortunately, due to the rather complicated form of variables $V_3$, $V^\alpha$, it is difficult to find a way to do this. 

Nevertheless, we can still derive the well known limit $\Lambda \longrightarrow 0$, ${\bf n} \longrightarrow 0$ (let us stress that one can also have $\Lambda = 0$ but ${\bf n} \neq 0$, see Subsection~\ref{sec:2.3}), which will allow to verify the correctness of our calculations. In this case the gauge group becomes the Poincar\'{e} group ${\rm SL}(2,\mathbbm{R}) \vartriangleright\!\!< \mathbbm{R}^{2,1}$, where the subgroup of translations $\mathbbm{R}^{2,1}$ is equivalent to the dual algebra $\mathfrak{sl}(2,\mathbbm{R})^*$. The semidirect product $\vartriangleright\!\!<$ means that the (double cover of the) Lorentz group ${\rm SL}(2,\mathbbm{R})$ is acting on $\mathfrak{sl}(2,\mathbbm{R})^*$ from the right, i.e.\! the product of two elements $g$ and $g^\prime$ has the form
\begin{align}\label{eq:21.06a}
g g^\prime = \mathfrak{u} (\mathbbm{1} + {\bf s})\, \mathfrak{u}^\prime (\mathbbm{1} + {\bf s}^\prime) = \mathfrak{u} \mathfrak{u}^\prime \left(\mathbbm{1} + (\mathfrak{u}^\prime)^{-1} {\bf s}\, \mathfrak{u}^\prime + {\bf s}^\prime\right)\,.
\end{align}
Consequently, $V_3$ and $V^\alpha$ significantly simplify to
\begin{align}\label{eq:21.07}
V_3 = \mu_3 v_3 - \tfrac{1}{4} \mu_0 v_0\,, \qquad V^\alpha = \mu_3 v^\alpha + \mu_0 \left(v_3 \eta^{\alpha 0} + \tfrac{1}{2} \epsilon^{0\alpha\beta} v_\beta\right)\,,
\end{align}
which is equivalent to $V_3 \mathbbm{1} + V^\alpha J_\alpha = \mathfrak{v}\, e^{\frac{1}{k} c_J \phi}$. Then (\ref{eq:21.06}) can be expressed via a total spatial derivative and integrated out to give the final particle Lagrangian
\begin{align}\label{eq:21.08}
L &= \frac{k}{2\pi} \int_\Gamma d\left<\partial_0(\mathfrak{v}\, e^{-\frac{1}{k} c_J \phi} \mathfrak{v}^{-1})\, \mathfrak{v}\, e^{\frac{1}{k} c_J \phi} \mathfrak{v}^{-1} {\bf x} + \frac{1}{k} c_P \phi\, \mathfrak{v}^{-1} \dot{\mathfrak{v}}\right> \nonumber\\
&= \kappa \left<\dot\Pi^{-1} \Pi\, {\bf x} + \frac{1}{\kappa} c_P \mathfrak{v}^{-1} \dot{\mathfrak{v}}\right>\,,
\end{align}
where we denote $\kappa \equiv k/2\pi$, and introduce new variables of the particle's momentum $\Pi \equiv \mathfrak{v}\, e^{c_J/\kappa} \mathfrak{v}^{-1}$ (see below) and position ${\bf x} \equiv \mathfrak{v}\, \bar{\bf s}\, \mathfrak{v}^{-1}$. Furthermore, to partially restrict the remaining gauge freedom we impose the natural condition $\gamma(\phi = 0) = 1$, which leads to the relations $\mathfrak{v} = \bar{\mathfrak{u}}$ and ${\bf r} = \bar{\mathfrak{u}}\, \bar{\bf s}\, \bar{\mathfrak{u}}^{-1}$, and hence we may set $\Pi = \bar{\mathfrak{u}}\, e^{c_J/\kappa} \bar{\mathfrak{u}}^{-1}$, ${\bf x} = \bar{\mathfrak{u}}\, \bar{\bf s}\, \bar{\mathfrak{u}}^{-1}$. The obtained result (\ref{eq:21.08}) obviously agrees with the previous derivations of the effective action \cite{Matschull:1997qy,Arzano:2014by} as well as symplectic form \cite{Meusburger:2003py}. 

Meanwhile, the holonomy of the connection $A_{\cal S} = \gamma d\gamma^{-1}$ around the boundary $\Gamma$ is given by the path-ordered exponential (with the counter-clockwise ordering)
\begin{align}\label{eq:21.09}
{\cal P}\, e^{\int_\Gamma A_{\cal S}} &= \gamma(\phi = 0)\, \gamma^{-1}(\phi = 2\pi) \nonumber\\ 
&= \Pi \left(\mathbbm{1} + \left({\rm Ad}(\Pi^{-1}) - 1\right) {\bf x} + {\rm Ad}(\Pi^{-1}) \left(\bar{\mathfrak{u}}\, \tfrac{1}{\kappa} c_P \bar{\mathfrak{u}}^{-1}\right)\right) \equiv 
\Pi \left(\mathbbm{1} + \tfrac{1}{\kappa} {\rm Ad}(\Pi^{-1})\, \Upsilon\right)\,,
\end{align}
where the adjoint action is ${\rm Ad}(\Pi)\, P_\alpha = \Pi\, P_\alpha \Pi^{-1}$. In this context a ${\rm SL}(2,\mathbbm{R})$ group element $\Pi$, conjugate to the rotation by $\frac{m}{\kappa} = 8\pi G m$, is naturally interpreted \cite{Matschull:1997qy,Meusburger:2005by} as momentum of the self-gravitating particle. Therefore, using the parametrization $\Pi := p_3 \mathbbm{1} + \frac{1}{\kappa} p^\alpha J_\alpha$ we obtain the deformed mass shell condition
\begin{align}\label{eq:21.10}
p_\alpha p^\alpha = 4\kappa^2 \sin^2\frac{m}{2\kappa}\,. 
\end{align}
The extended momentum space, which is the ${\rm SL}(2,\mathbbm{R})$ group, as a manifold is the three-dimensional anti-de Sitter space, determined by the constraint on coordinates $p_3^2 + p_\alpha p^\alpha/(4\kappa^2) = 1$. On the other hand, $\Upsilon = j^\alpha P_\alpha$ introduced above can be shown \cite{Matschull:1997qy,Meusburger:2003py} to be the particle's (generalized) angular momentum, with deformed components
\begin{align}\label{eq:21.11}
j^\alpha = p_3\, \epsilon^{\alpha}_{\ \beta\gamma} x^\beta p^\gamma + 
\frac{1}{2\kappa} \left(x^\alpha p_\beta p^\beta - x_\beta p^\beta p^\alpha\right) + s\, \hat p^\alpha\,.
\end{align}
Together with $\Pi$ it satisfies the relation $p_\alpha j^\alpha = 4\kappa^2 \sin^2\frac{m}{2\kappa}\, s$ instead of the usual $p_\alpha j^\alpha = m s$ (which is valid for ${\bf p}$ and ${\bf j}$ in the starting action (\ref{eq:20.04})). 

We also note that the effective Lagrangian (\ref{eq:21.08}) can be expressed in the concise form
\begin{align}\label{eq:21.15}
L = \left<\dot{\bar{\mathfrak{u}}} \bar{\mathfrak{u}}^{-1} \Upsilon\right>\,.
\end{align}
For vanishing spin $s = 0$, evaluating the scalar product in the Lagrangian and treating the mass shell condition (\ref{eq:21.10}) as a constraint, we can write the corresponding action in terms of coordinates $x^\alpha$ and $p^\alpha$ as
\begin{align}\label{eq:21.12}
S &= \int\! dt\ \left(\left(-p_3 \dot p_\alpha + \dot p_3 p_\alpha - 
\frac{1}{2\kappa}\, \epsilon_{\alpha\beta\gamma} \dot p^\beta p^\gamma\right) x^\alpha - \frac{\lambda}{2} \left(p_\alpha p^\alpha - 4\kappa^2 \sin^2\frac{m}{2\kappa}\right)\right)\,,
\end{align}
where $\frac{\lambda}{2}$ is the Lagrange multiplier (and $p_3 = \sqrt{1 - p_\alpha p^\alpha/(4\kappa^2)}$). As one can see, in the no-gravity limit $\kappa \longrightarrow \infty$ it reduces to the free particle action. Surprisingly, after some calculations, (\ref{eq:21.12}) itself gives the equations of motion
\begin{align}\label{eq:21.13}
\dot x^\alpha = \lambda \cos\frac{m}{2\kappa}\, p^\alpha\,, \qquad 
\dot p^\alpha = 0\,,
\end{align}
which are the same (up to the rescaling of $\lambda$) as for a free particle \cite{Matschull:1997qy,Kowalski:2014dy}. What is actually modified by the gravitational field is the momentum mass shell. These conclusions remain valid in the spinning case, when the on-shell action is
\begin{align}\label{eq:21.14}
S = \int\! dt\ \left(\left(-p_3 \dot p_\alpha + \dot p_3 p_\alpha - 
\frac{1}{2\kappa}\, \epsilon_{\alpha\beta\gamma} \dot p^\beta p^\gamma\right) x^\alpha + \frac{s}{2}\, \epsilon_{0\alpha}^{\ \ \beta} \dot \Lambda^\alpha_{\ \gamma}(-\bar{\bf u}) \Lambda^\gamma_{\ \beta}(\bar{\bf u})\right)
\end{align}
where $\Lambda^\alpha_{\ \beta}(\bar{\bf u})$ are matrix elements of the adjoint representation of the Lorentz group, defined via $\Lambda^\alpha_{\ \beta}(\bar{\bf u}) J_\alpha := \bar{\mathfrak{u}} J_\beta \bar{\mathfrak{u}}^{-1}$, while we implicitly have $p^\alpha = 2\kappa\, \Lambda^\alpha_{\ 0}(\bar{\bf u}) \sin\frac{m}{2\kappa}$ and $x^\alpha = \Lambda^\alpha_{\ \beta}(\bar{\bf u})\, \bar s^\beta$. Variating (\ref{eq:21.14}) with respect to $\Lambda^\alpha_{\ \beta}(\bar{\bf u})$, it can independently be shown (similarly as it is done for a free particle \cite{Balachandran:2015gs}) that $\dot j_\alpha = 0$. 

Finally, let us briefly discuss the case of a massless particle \cite{Matschull:1997qy}, which is usually neglected in the Chern-Simons formulation. In order to correctly include such a particle in the action (\ref{eq:20.04}) we have to replace $c_0$ with a lightlike algebra element, so that $p_\alpha p^\alpha = 0$ for $h c_J h^{-1} = p^\alpha J_\alpha$ and $[c_J,c_P] = 0$, e.g.\! $c_J = e\, (J_0 \pm J_2)$, $c_P = s\, (P_0 \pm P_2)$, where $e$ is the massless particle's energy. As one can verify, the derivation of (\ref{eq:21.08}) (obviously, we do not need to start it from $\Lambda > 0$) is unaffected by this change and we arrive at the effective Lagrangian of the same form as in the massive case. The off-shell particle action in terms of coordinates can be simply obtained by taking the limit $m \longrightarrow 0$ of (\ref{eq:21.12}). However, now $\Pi := p_3 \mathbbm{1} + \frac{1}{\kappa} p^\alpha J_\alpha$ on shell is conjugate to the null rotation by $8\pi G e$, while $p^\alpha = e\, (\Lambda^\alpha_{\ 0}(\bar{\bf u}) + \Lambda^\alpha_{\ 1}(\bar{\bf u}))$.

\subsection{The reciprocal group contraction} \label{sec:2.2}
From the point of view of the (local) gauge group factorization (\ref{eq:20.12}), curved manifold of the ${\rm AN}_{\bf n}(2)$ component group in the limit $\Lambda \longrightarrow 0$ is being flattened to the $\mathbbm{R}^{2,1}$ group. On the other hand, one might theoretically consider the opposite contraction of the de Sitter group, such that the Lorentz component ${\rm SL}(2,\mathbbm{R})$ is flattened out and we obtain the group with the semidirect product structure $\mathbbm{R}^{2,1} >\!\!\vartriangleleft {\rm AN}_{\bf n}(2)$, equivalent to $\mathfrak{an}_{\bf n}(2)^* >\!\!\vartriangleleft {\rm AN}_{\bf n}(2)$ (with the left action of ${\rm AN}_{\bf n}(2)$ on $\mathfrak{an}_{\bf n}(2)^*$). This should lead to the particle model with the ${\rm AN}_{\bf n}(2)$ momentum space instead of ${\rm SL}(2,\mathbbm{R})$ and might be connected with the $\kappa$-Poincar\'{e} symmetry algebra. The proper way to accomplish this task, which we did not completely explain in \cite{Kowalski:2014dy}, is by rescaling the de Sitter group generators and coordinates to
\begin{align}\label{eq:22.01}
\tilde J_\alpha := \sqrt{\Lambda}\, J_\alpha\,, \quad \tilde S_\alpha := \sqrt{\Lambda}^{-1} S_\alpha\,, \qquad \tilde u^\alpha := \sqrt{\Lambda}^{-1} u^\alpha\,, \quad \tilde s^\alpha := \sqrt{\Lambda}\, s^\alpha
\end{align}
(below we will skip the tildes over coordinates) and subsequently taking the limit $\Lambda \longrightarrow 0$. After such a contraction the deformation vector effectively reduces to ${\bf n} = (1,0,0)$, while the brackets (\ref{eq:A1.07}) (with $\Lambda > 0$) become
\begin{align}\label{eq:22.02}
[\tilde J_\alpha,\tilde J_\beta] = 0\,, \qquad 
[\tilde J_\alpha,\tilde S_\beta] = \eta_{\beta 0} \tilde J_\alpha - 
\eta_{\alpha\beta} \tilde J_0\,, \qquad 
[\tilde S_\alpha,\tilde S_\beta] = \eta_{\alpha 0} \tilde S_\beta - 
\eta_{\beta 0} \tilde S_\alpha\,,
\end{align}
which indeed describe the Lie algebra corresponding to $\mathfrak{an}_{\bf n}(2)^* >\!\!\vartriangleleft {\rm AN}_{\bf n}(2)$. The product of two group elements $g$ and $g^\prime$ has the form
\begin{align}\label{eq:22.02a}
g g^\prime = (\mathbbm{1} + {\bf u}) \mathfrak{s}\, (\mathbbm{1} + {\bf u}^\prime) \mathfrak{s}^\prime = \left(\mathbbm{1} + {\bf u} + \mathfrak{s}\, {\bf u}^\prime \mathfrak{s}^{-1}\right) \mathfrak{s} \mathfrak{s}^\prime\,.
\end{align}
The scalar product (\ref{eq:20.11}) on the algebra (\ref{eq:22.02}) remains unchanged, due to the fact that we rescaled both subalgebras rather than just $\mathfrak{sl}(2,\mathbbm{R})$. The physical meaning of the Chern-Simons theory with the obtained new gauge group will become clear below. 

Let us first observe that the algebra (\ref{eq:22.02}) bears a certain resemblance to algebra of the three-dimensional Carroll group. This group is the contraction of the (three-dimensional) Poincar\'{e} group defined in the limit of vanishing speed of light \cite{Bacry:1968ps} but can also be seen as a subgroup of the de Sitter group, treated as a so-called Bargmann group \cite{Duval:2014ce}. The Carroll algebra is given by
\begin{align}\label{eq:22.03}
[M,K_a] &= \epsilon_{0ab} K^b\,, & [K_a,K_b] &= 0\,, & 
[M,T_a] &= \epsilon_{0ab} T^b\,, & [K_a,T_b] &= \delta_{ab} T_0\,, \nonumber\\ 
[M,T_0] &= 0\,, & [K_a,T_0] &= 0\,, & [T_0,T_a] &= 0\,, & 
[T_a,T_b] &= 0\,,
\end{align}
where $M$, $K_a$, $T_a$, $T_0$, $a = 1,2$ are, respectively, the generators of rotations, Carrollian boosts and translations in space and time. Carrollian boosts are acting only in the temporal direction and the differences between (\ref{eq:22.03}) and the Poincar\'{e} algebra are associated with them. To facilitate a comparison of the Carroll algebra structure with the brackets (\ref{eq:22.02}) we may rewrite the latter in the form
\begin{align}\label{eq:22.04}
[\tilde M,\tilde K_a] &= \tilde K_a\,, & [\tilde K_a,\tilde K_b] &= 0\,, & 
[\tilde M,\tilde T_a] &= -\tilde T_a\,, & [\tilde K_a,\tilde T_b] &= \delta_{ab} \tilde T_0\,, \nonumber\\ 
[\tilde M,\tilde T_0] &= 0\,, & [\tilde K_a,\tilde T_0] &= 0\,, & 
[\tilde T_0,\tilde T_a] &= 0\,, & [\tilde T_a,\tilde T_b] &= 0\,,
\end{align}
where we denoted $\tilde M \equiv \tilde S_0$, $\tilde K_a \equiv \tilde S_a$, $\tilde T_a \equiv \tilde J_a$ and $\tilde T_0 \equiv -\tilde J_0$. Looking at (\ref{eq:22.03}) we observe that $\tilde J_\alpha$'s, $\tilde S_0$ and $\tilde S_a$'s occupy the respective positions of the generators of translations, rotations and Carrollian boosts but with the altered first and third bracket. As we will see, this strong similarity between the algebras (\ref{eq:22.02}) and (\ref{eq:22.03}) manifests itself at the level of particle dynamics. 

We note that the effective symplectic form for particles coupled to the Chern-Simons theory with a gauge group of the form $G \vartriangleright\!\!< \mathfrak{g}^*$, where $G$ is an arbitrary Lie group, was calculated in \cite{Meusburger:2006pe}. Starting from (\ref{eq:21.06}), we will now finish the derivation of the particle action in the case of the considered group $\mathfrak{an}_{\bf n}(2)^* >\!\!\vartriangleleft {\rm AN}_{\bf n}(2)$. It simplifies the sewing condition (\ref{eq:20.09}) to $V_3 = 1$ and
\begin{align}\label{eq:22.05}
V_0 = \mu_0 + v_0 + (\sigma_3 + \tfrac{1}{2} \sigma_0)^2 (\bar s_3 - \tfrac{1}{2} \bar s_0)\, v^a \bar s_a\,, \qquad 
V^a = (\sigma_3 + \tfrac{1}{2} \sigma_0)^2 (\bar s_3 - \tfrac{1}{2} \bar s_0)^2 v^a\,,
\end{align}
which, as can be shown using formulae from the Appendix, is equivalent to the relation 
\begin{align}\label{eq:22.06}
V_3 \mathbbm{1} + V^\alpha \tilde J_\alpha = \bar{\mathfrak{s}}\, e^{-\frac{1}{k} c_{\tilde S} \phi} {\bf v}\, e^{\frac{1}{k} c_{\tilde S} \phi} \bar{\mathfrak{s}}^{-1} + e^{\frac{1}{k} c_{\tilde J} \phi}\,.
\end{align}
Hence we find that the particle Lagrangian (\ref{eq:21.06}) becomes
\begin{align}\label{eq:22.07}
L &= \frac{k}{2\pi} \int_\Gamma d\left<\partial_0(\bar{\mathfrak{s}}\, e^{-\frac{1}{k} c_{\tilde S} \phi} \bar{\mathfrak{s}}^{-1} {\bf x}\, \bar{\mathfrak{s}})\, e^{\frac{1}{k} c_{\tilde S} \phi} \bar{\mathfrak{s}}^{-1} + \frac{1}{k} c_{\tilde J} \phi\, \bar{\mathfrak{s}}^{-1} \dot{\bar{\mathfrak{s}}}\right> \nonumber\\
&= \kappa \left<\Pi \dot\Pi^{-1} {\bf x} + \frac{1}{\kappa} c_{\tilde J} \bar{\mathfrak{s}}^{-1} \dot{\bar{\mathfrak{s}}}\right>\,,
\end{align}
where $\Pi \equiv \bar{\mathfrak{s}}\, e^{c_{\tilde S}/\kappa} \bar{\mathfrak{s}}^{-1}$ and ${\bf x} \equiv \bar{\mathfrak{s}}\, {\bf v}\, \bar{\mathfrak{s}}^{-1}$. We can again fix the gauge via the condition $\gamma(\phi = 0) = 1$, equivalent to $\mathfrak{r} = \bar{\mathfrak{s}}$, ${\bf v} = \bar{\mathfrak{s}}^{-1} \bar{\bf u}\, \bar{\mathfrak{s}}$, which allows us to write ${\bf x} = \bar{\bf u}$. The Lagrangian (\ref{eq:22.07}) is the confirmation of our findings from \cite{Kowalski:2014dy} but here we arrive at this result in a more rigorous manner. Let us also note that, despite some differences, the first lines of (\ref{eq:22.07}) and (\ref{eq:21.08}) lead to the almost identical form of the final expression. More precisely, (\ref{eq:22.07}) corresponds to the counterpart of (\ref{eq:21.08}) with the reverse factorization (\ref{eq:A1.11}) of gauge group elements. 

Calculating the holonomy of $A_{\cal S}$ around $\Gamma$ we then obtain
\begin{align}\label{eq:22.08}
{\cal P}\, e^{\int_\Gamma A_{\cal S}} = \gamma(\phi = 0)\, \gamma^{-1}(\phi = 2\pi) = \left(\mathbbm{1} + \left(1 - {\rm Ad}(\Pi)\right) {\bf x} + \tfrac{1}{\kappa} c_{\tilde J}\right) \Pi\,.
\end{align}
which may be compared with (\ref{eq:21.09}). By analogy with the latter case, we presume that a group element $\Pi \in {\rm AN}_{\bf n}(2)$ is actually the particle's momentum and ${\bf x} \in \mathbbm{R}^{2,1}$ its position. Indeed, after the rescaling (\ref{eq:22.01}) the mass constant $\tilde m \equiv \sqrt{\Lambda}^{-1} m$ acquires the dimension of mass times length (i.e.\! angular momentum), while the spin constant $\tilde s \equiv \sqrt{\Lambda}\, s$ the dimension of mass (i.e.\! momentum). In this sense mass and spin are exchanged and therefore we will denote $s\, \tilde J_0 := c_{\tilde J}$ and $m\, \tilde S_0 := c_{\tilde S}$. As expected due to the constraints (\ref{eq:20.06}), it can also be shown that a similar exchange occurs for the spin connection and dreibein in the Cartan connection (\ref{eq:20.03}) (now written in terms of the generators $\tilde J_\alpha$ and $\tilde S_\alpha$). Therefore, the theory with the gauge group considered in this Subsection can be seen as related to the Poincar\'{e} case from the previous Subsection via a kind of the reciprocity (or duality) map. However, since the new dreibein and spin connection do not lead to the standard expressions (\ref{eq:20.03a}) for the Riemann curvature and torsion, such a theory is actually a modification of general relativity. 

The extended momentum manifold ${\rm AN}_{\bf n}(2)$ is the elliptic de Sitter space \cite{Kowalski:2003de}. In the context of the $\kappa$-Poincar\'{e} algebra, ${\rm AN}_{\bf n}(2)$ is often considered in the exponential parametrization, e.g.\! with the following ordering
\begin{align}\label{eq:22.09}
\Pi := e^{p^a/\kappa\, \tilde S_a} e^{p^0/\kappa\, \tilde S_0}\,, \qquad 
\bar{\mathfrak{s}} := e^{\xi^a \tilde S_a} e^{\xi^0 \tilde S_0}\,.
\end{align}
It is connected with the parametrization $\mathfrak{s} = s_3 \mathbbm{1} + s^\alpha \tilde S_\alpha$ by the relations $\xi^0 = 2 \log(s_3 + \frac{1}{2} s^0)$ and $\xi^a = (s_3 + \frac{1}{2} s^0)\, s^a$. However, it turns out that coordinates $p^\alpha$ are constrained to \cite{Kowalski:2014dy}
\begin{align}\label{eq:22.10}
p^0 = m\,, \qquad p^a = \kappa \left(1 - e^{\frac{m}{\kappa}}\right) \xi^a\,,
\end{align}
which means that the particle's energy $p^0$ always has to be equal to the rest mass. This is a characteristic feature for particles with the Carroll group symmetry since in the Carrollian (or ultralocal) limit lightcones in spacetime are shrunk into spacelike worldlines, which can equivalently be seen as null geodesics in one dimension higher \cite{Duval:2014ce}. By analogy with (\ref{eq:21.09}), we also call the quantity
\begin{align}\label{eq:22.11}
\Upsilon \equiv \kappa\, ({\rm Ad}(\Pi^{-1}) - 1)\, {\bf x} + c_{\tilde J} = \left(-x_b p^b + s\right) \tilde J_0 + x^a p_0 \tilde J_a
\end{align}
the particle's (quasi-)angular momentum. It satisfies the standard condition $p^\alpha j_\alpha = m s$ but in the limit $\kappa \longrightarrow \infty$ its components become the expressions $j^0 = -x_a p^a + s$, $j^a = p_0 x^a$, which are different than in the ordinary situation. 

The Lagrangian (\ref{eq:22.07}) can concisely be written as
\begin{align}\label{eq:22.12}
L = \left<\dot{\bar{\mathfrak{s}}} \bar{\mathfrak{s}}^{-1} \Upsilon\right>\,.
\end{align}
However, since there is the relation $\bar{\mathfrak{s}} \tilde J_0 \bar{\mathfrak{s}}^{-1} = \tilde J_0$, calculating a variation of the spin term we obtain the total time derivative $\delta \left<c_{\tilde J} \bar{\mathfrak{s}}^{-1} \dot{\bar{\mathfrak{s}}}\right> = \partial_0 \left<c_{\tilde J} \varepsilon\right>$, where $\delta \bar{\mathfrak{s}} = \varepsilon\, \bar{\mathfrak{s}}$, $\varepsilon \in \mathfrak{an}_{\bf n}(2)$. Since it does not contribute to the equations of motion, let us now restrict to the spinless case. The action corresponding to the Lagrangian (\ref{eq:22.07}), with the mass shell constraint (\ref{eq:22.10}) included, in coordinates is given by
\begin{align}\label{eq:22.13}
S = \int\! dt\ \left(\dot x^0 p_0 + \dot x^a p_a + \kappa^{-1} x^a p_a \dot p_0 - \frac{\lambda}{2} \left(p_0^2 - m^2\right)\right)\,.
\end{align}
Infinitesimal symmetries of this action are described in \cite{Kowalski:2014dy}. Without the constraint it would be the off-shell action of a particle with the $\kappa$-Poincar\'{e} symmetries \cite{Imilkowska:2006dy}. On the other hand, in the $\kappa \longrightarrow \infty$ limit it becomes the action of a free Carroll particle \cite{Bergshoeff:2014ds} in three dimensions. Furthermore, (\ref{eq:22.13}) leads to the equations of motion (on the mass shell)
\begin{align}\label{eq:22.14}
\dot x^0 = \lambda\, m\,, \qquad \dot x^a = 0\,, \qquad \dot p^\alpha = 0\,,
\end{align}
which are actually identical to the ones of a Carroll particle. Taking everything into account, one can say that the action (\ref{eq:22.13}) describes a $\kappa$-deformed Carroll particle, although the underlying gauge group is not a $\kappa$-deformation of the Carroll group.

\subsection{Anti-de Sitter group and the lightlike deformation} \label{sec:2.3}
Another case to analyze is the Lagrangian (\ref{eq:20.14}) with $\Lambda < 0$ and the anti-de Sitter gauge group ${\rm SL}(2,\mathbbm{R}) \times {\rm SL}(2,\mathbbm{R})$. Choosing the deformation vector ${\bf n} = (0,0,\sqrt{|\Lambda|})$, we factorize a constant $\alpha$ in the sewing condition (\ref{eq:20.09}) in the same way as in (\ref{eq:21.03}) (but now we have $r_3 + \tfrac{1}{2} \sqrt{|\Lambda|}\, r_2 > 0$) and write the group element of a particle as
\begin{align}\label{eq:23.01}
e^{\frac{1}{k} c_0 \phi} = e^{\frac{1}{k} c_P \phi} e^{\frac{1}{k} c_J \phi} = \mathfrak{r}_c \mathfrak{v}_c\,.
\end{align}
However, in this case the second line of (\ref{eq:21.02}) becomes 
\begin{align}\label{eq:23.02}
\sigma_3 = \cos\frac{\sqrt{|\Lambda|}\, s \phi}{2k}\,, \qquad \sigma_0 = \frac{2}{\sqrt{|\Lambda|}}\, \sin\frac{\sqrt{|\Lambda|}\, s \phi}{2k}\,, 
\end{align}
while (\ref{eq:23.01}) acquires the nontrivial form
\begin{align}\label{eq:23.03}
\mathfrak{v}_c &= \frac{1}{\sqrt{\sigma_3^2 + \frac{\Lambda}{4} \sigma_0^2}} \left(\mu_3 \sigma_3 \mathbbm{1} + \mu_0 \sigma_3 J_0 - \sqrt{|\Lambda|}\, \mu_3 \sigma_0 J_1 - \tfrac{1}{2} \sqrt{|\Lambda|}\, \mu_0 \sigma_0 J_2\right)\,, \nonumber\\
\mathfrak{r}_c &= \frac{1}{\sqrt{\sigma_3^2 + \frac{\Lambda}{4} \sigma_0^2}} \left(\sigma_3^2 \mathbbm{1} + \sigma_3 \sigma_0 S_0 - \tfrac{1}{2} \sqrt{|\Lambda|}\, \sigma_0^2 S_2\right)\,,
\end{align}
and the corresponding factorization condition $\sigma_3^2 + \frac{\Lambda}{4} \sigma_0^2 = \cos(\sqrt{|\Lambda|}\, s \phi/k) > 0$ is not always satisfied. We subsequently calculate that the counterparts of the formulae (\ref{eq:21.04}-\ref{eq:21.05}) are given by
\begin{align}\label{eq:23.04}
\mathfrak{u}^{-1} = (V_3 \mathbbm{1} + V^\alpha J_\alpha)\, \bar{\mathfrak{u}}^{-1}\,, \quad 
V_3 &= \frac{1}{N_V} \left(U_3 \bar s_3 + \tfrac{1}{2} \sqrt{|\Lambda|} (U_3 \bar s_2 + \epsilon_{\alpha\beta 2} U^\alpha \bar s^\beta)\right)\,, \nonumber\\
V^\alpha &= \frac{1}{N_V} \left((\bar s_3 - \tfrac{1}{2} \sqrt{|\Lambda|}\, \bar s_2) U^\alpha + \sqrt{|\Lambda|}\, U^\beta \bar s_\beta \eta^{\alpha 2}\right)\,, \nonumber\\
N_V &\equiv \sqrt{(\bar s_3 + \tfrac{1}{2} \sqrt{|\Lambda|}\, \bar s_2)^2 + \tfrac{1}{4} \Lambda\, \bar s^\alpha \bar s_\alpha (U_0^2 + U_1^2)} \nonumber\\
&\overline{+ \sqrt{|\Lambda|} (\bar s_3 + \tfrac{1}{2} \sqrt{|\Lambda|}\, \bar s_2) \left(\tfrac{1}{2} (U^\alpha \bar s_\alpha U_2 - U^\alpha U_\alpha \bar s_2) + U_3 \epsilon_{\alpha\beta 2} U^\alpha \bar s^\beta\right)}\,,
\end{align}
which exists when $N_V^2 > 0$, and
\begin{align}\label{eq:23.05}
U_3 \mathbbm{1} + U^\alpha J_\alpha = (W_3 \mathbbm{1} + W^\alpha J_\alpha)\, \mathfrak{v}_c\,, \quad 
W_3 &= \frac{1}{N_W \sqrt{\sigma_3^2 + \frac{\Lambda}{4} \sigma_0^2}} \left((\sigma_3^2 + \tfrac{\Lambda}{4} \sigma_0^2) v_3 - \tfrac{1}{2} \sqrt{|\Lambda|}\, \sigma_3 \sigma_0 v_1\right)\,, \nonumber\\
W^\alpha &= \frac{1}{N_W \sqrt{\sigma_3^2 + \frac{\Lambda}{4} \sigma_0^2}} \left(v^\alpha - \sqrt{|\Lambda|}\, \sigma_0 (\sigma_3 v_0 - \tfrac{1}{2} \sqrt{|\Lambda|}\, \sigma_0 v_2) \eta^{\alpha 2}\right)\,, \nonumber\\
N_W &\equiv \sqrt{\sigma_3^2 + \tfrac{\Lambda}{4} \sigma_0^2 - \sqrt{|\Lambda|}\, \sigma_3 \sigma_0 (v_3 v_1 + \tfrac{1}{2} v_0 v_2)}\,,
\end{align}
which has to satisfy $N_W^2 > 0$. For brevity we do not present the explicit expressions for $U_3$, $U^\alpha$. However, similarly as it is for $\Lambda > 0$, we do not know how to perform the final integration in the particle Lagrangian (\ref{eq:21.06}) after substituting (\ref{eq:23.04}-\ref{eq:23.05}). In the limit $\Lambda \longrightarrow 0$ we obviously recover the sewing condition for the Poincar\'{e} gauge group (\ref{eq:21.07}) and the corresponding ultimate Lagrangian (\ref{eq:21.08}). On the other hand, although in the case of the anti-de Sitter group one can also define the counterpart of the group contraction from the previous Subsection, the obtained Lagrangian turns out to be divergent. This result seems to be associated with the term proportional to $J_1$ in $\mathfrak{v}_c$. 

Lastly, we may take $\Lambda = 0$ but with a non-zero, lightlike deformation vector, e.g.\! ${\bf n} = (q,0,q)$, $q \in \mathbbm{R}$. We find that the sewing condition is then given by complicated expressions analogous to (\ref{eq:23.04}-\ref{eq:23.05}) and therefore we do not show them here. It might seem that the effective action in such a case should be equivalent to the one for the Poincar\'{e} group in the standard form ${\rm SL}(2,\mathbbm{R}) \vartriangleright\!\!< \mathbbm{R}^{2,1}$ but (\ref{eq:21.08}) is recovered in the limit $q \longrightarrow 0$, which reflects the fact that $q$ is an extra parameter. The situation of the reciprocal group contraction is the same as above for $\Lambda < 0$.

\section{Multiple particles} \label{sec:3.0}
We will now generalize our derivation of the effective action to multiple particles. In the Subsections below we concentrate on these gauge groups for which we have managed to obtain the final form of the single particle Lagrangian but what we do previously is valid for any $\Lambda$. The starting point is the Chern-Simons action for a system of $n$ particles coupled to gravity \cite{Meusburger:2009gy}, given by the straightforward counterpart to the single particle case (\ref{eq:20.04})
\begin{align}\label{eq:30.01}
S = \int dt L\,, \quad L_n &= \frac{k}{4\pi} \int_{\cal S} \left<\dot A_{\cal S} \wedge A_{\cal S}\right> - 
\sum_{i=1}^n \left<c_{(i)} h_i^{-1} \dot h_i\right> \nonumber\\ 
&+ \int_{\cal S} \left<A_0 \left(\frac{k}{2\pi} F_{\cal S} - 
\sum_{i=1}^n h_i c_{(i)} h_i^{-1} \delta^2(\vec{x} - \vec{x}_{(i)})\, dx^1 \wedge dx^2\right)\right>\,.
\end{align}
where particles are labelled by $i = 1,\ldots,n$ and appear as punctures at points $\vec{x}_{(i)}$ of a spatial slice ${\cal S}$ of genus $0$, while their masses and spins are encoded in the algebra elements $c_{(i)} = m_{(i)} J_0 + s_{(i)} P_0$, and momenta and angular momenta determined by the group elements $h_i$. The topology of ${\cal S}$ can be either open or closed (for $\Lambda \leq 0$ the latter is possible when $n \geq 3$ \cite{Deser:1984tc}). In the open case we also should impose the appropriate boundary conditions at spatial infinity, which are given by the requirement that spacetime is asymptotically conical, corresponding to a single effective particle \cite{Matschull:2001ty}. It can elegantly be done \cite{Meusburger:2005by,Meusburger:2006pe} by treating ${\cal S}$ as a topological sphere on which the infinity is represented by the boundary of a removed disc, and then shrinking this boundary into a special additional puncture, which carries the total mass and spin of the system. Nevertheless, for simplicity we consider here the reduced setting, assuming that the boundary conditions are already satisfied. 

To solve the constraint on $F_{\cal S}$ from the second line of (\ref{eq:30.01}) it is convenient to decompose ${\cal S}$ in the manner \cite{Alekseev:1995se,Meusburger:2005by} that generalizes what we did in the previous Section. We first choose a point on ${\cal S}$ far away from the punctures and starting from it draw a separate loop around each of them, dividing ${\cal S}$ into $n$ disjoint particle regions ${\cal D}_i$ and the asymptotic empty region ${\cal E}$ with the boundary $\Gamma$. Similarly as before, every region ${\cal D}_i$ can be deformed into a punctured disc, with polar coordinates $\rho_i \in (0,1]$, $\phi_i \in [0,2\pi]$, where the connection $A_{\cal S}$ has the form
\begin{align}\label{eq:30.02}
A_{\cal S}^{({\cal D}_i)} = \bar\gamma_i\, \frac{c_{(i)}}{k} d\phi_i\, 
\bar\gamma_i^{-1} + \bar\gamma_i d\bar\gamma_i^{-1}\,, \quad 
\bar\gamma_i(\rho_i = 0) = h_i\,.
\end{align}
Meanwhile, the empty region ${\cal E}$ (where $A_{\cal S}$ is given by (\ref{eq:20.07})) can be seen as a $n$-sided polygon whose edges $\Gamma_i$ ($\Gamma = \bigcup_i \Gamma_i$) correspond to the boundaries of discs ${\cal D}_i$. At the $i$'th vertex of the polygon the endpoint $\phi_i = 2\pi$ of the incoming edge $\Gamma_i$ coincides with the endpoint $\phi_{i+1} = 0$ of the outgoing edge $\Gamma_{i+1}$. However, on every $\Gamma_i$ we have an independent sewing condition $A_{\cal S}^{({\cal D}_i)}|_{\Gamma_i} = A_{\cal S}^{({\cal E})}|_{\Gamma_i}$ and therefore we can apply the same methods as for a single particle. 

\subsection{The Poincar\'{e} case} \label{sec:3.1}
Let us first restrict to the Poincar\'{e} gauge group, with $\Lambda = 0$ and ${\bf n} = 0$. Following Section~\ref{sec:2.0}, for each particle we derive the effective Lagrangian of the form (\ref{eq:21.08}), i.e. (after evaluating the scalar product)
\begin{align}\label{eq:31.01}
L_{(i)} = \kappa \left(\dot\Pi_i^{-1} \Pi_i\right)_\alpha \left({\bf x}_i\right)^\alpha + 
s_{(i)} \left(\mathfrak{v}_i^{-1} \dot{\mathfrak{v}}_i\right)_0\,,
\end{align}
with momentum $\Pi_i = \mathfrak{v}_i e^{\frac{1}{\kappa} m_{(i)} J_0} \mathfrak{v}_i^{-1}$ and position ${\bf x}_i = \mathfrak{v}_i \bar{\mathfrak{s}}_i \mathfrak{v}_i^{-1}$. Moreover, we have to ensure the continuity of $\gamma$ at all vertices of ${\cal E}$ except $i = 1$ (where $\gamma$ has a jump, analogously to (\ref{eq:20.09})), imposing the conditions $\gamma(\phi_{i+1} = 0) = \gamma(\phi_i = 2\pi)$, $i < n$. Similarly as in the previous Section, we may also fix the gauge at the first vertex via $\gamma(\phi_1 = 0) = 1$. Together this leads to the sequence of relations
\begin{align}\label{eq:31.02}
\mathfrak{v}_1 \bar{\mathfrak{u}}_1^{-1} &= 1\,, & 
\mathfrak{v}_2 \bar{\mathfrak{u}}_2^{-1} &= \Pi_1\,, & 
\mathfrak{v}_3 \bar{\mathfrak{u}}_3^{-1} &= \Pi_1 \Pi_2\,,\,\ldots\,, \nonumber\\ 
{\bf r}_1 &= {\bf x}_1\,, & 
{\bf r}_2 &= \Pi_1 {\bf x}_2 \Pi_1^{-1} + \tfrac{1}{\kappa} \Upsilon_1\,, & 
{\bf r}_3 &= \Pi_1 \Pi_2 {\bf x}_3 \Pi_2^{-1} \Pi_1^{-1} + \Pi_1 \tfrac{1}{\kappa} \Upsilon_2 \Pi_1^{-1} + \tfrac{1}{\kappa} \Upsilon_1\,,\,\ldots\,,
\end{align}
where now $\Pi_i \equiv \bar{\mathfrak{u}}_i e^{\frac{1}{\kappa} m_{(i)} J_0} \bar{\mathfrak{u}}_i^{-1}$, ${\bf x}_i \equiv \bar{\mathfrak{u}}_i \bar{\mathfrak{s}}_i \bar{\mathfrak{u}}_i^{-1}$ and angular momentum $\Upsilon_i = \kappa\, (1 - {\rm Ad}(\Pi_i))\, {\bf x}_i + \bar{\mathfrak{u}}_i s_{(i)} P_0 \bar{\mathfrak{u}}_i^{-1}$. 

Substituting the conditions (\ref{eq:31.02}) into individual Lagrangians (\ref{eq:31.01}), we choose to eliminate variables $\mathfrak{v}_i$ in favour of $\bar{\mathfrak{u}}_i$. Performing the summation over all particles we then obtain the effective $n$-particle Lagrangian, which can be written as
\begin{align}\label{eq:31.03}
L_n &= \sum_{i=1}^n \left(\dot{\bar{\mathfrak{u}}}_i \bar{\mathfrak{u}}_i^{-1} - \partial_0(\Pi_{i-1}^{-1} \ldots \Pi_1^{-1})\, \Pi_1 \ldots \Pi_{i-1} \right)_\alpha \left(\Upsilon_i\right)^\alpha \nonumber\\ 
&= \sum_{i=1}^n \left(\kappa \left(\dot\Pi_i^{-1} \Pi_i\right)_\alpha \left({\bf x}_i\right)^\alpha + s_{(i)} \left(\bar{\mathfrak{u}}_i^{-1} \dot{\bar{\mathfrak{u}}}_i\right)_0 - \left(\partial_0(\Pi_{i-1}^{-1} \ldots \Pi_1^{-1})\, \Pi_1 \ldots \Pi_{i-1} \right)_\alpha \left(\Upsilon_i\right)^\alpha\right)\,.
\end{align}
In particular, in the 3-particle case the explicit expression for $L_n$ is (here we arrange it in a different way)
\begin{align}\label{eq:31.04}
L_3 = \sum_{i=1}^3 \left(\kappa \left(\dot\Pi_i^{-1} \Pi_i\right)_\alpha \left({\bf x}_i\right)^\alpha + s_{(i)} \left(\bar{\mathfrak{u}}_i^{-1} \dot{\bar{\mathfrak{u}}}_i\right)_0\right) - \left(\dot\Pi_1^{-1} \Pi_1\right)_\alpha \left(\Upsilon_2 + \Pi_2 \Upsilon_3 \Pi_2^{-1}\right)^\alpha - \left(\dot\Pi_2^{-1} \Pi_2\right)_\alpha \left(\Upsilon_3\right)^\alpha\,.
\end{align}
The Lagrangian (\ref{eq:31.03}) agrees with the corresponding symplectic form \cite{Meusburger:2005by}. As one can observe, it describes a collection of self-gravitating particles whose angular momentum $\Upsilon_i$ is coupling to the total momentum of preceding particles (i.e.\! the ones labelled by $j$'s smaller than a given $i \leq n$). Furthermore, the terms proportional to $\Upsilon_i$ depend on the order of particle labels. 

To verify the composition rule for such group-valued momenta and angular momenta we note that the holonomy of $A_{\cal S}$ along an edge $\Gamma_i$ is given by $\gamma(\phi_i = 0)\, \gamma^{-1}(\phi_i = 2\pi)$ (similarly to (\ref{eq:21.09})), and hence for the holonomy circumventing $j \leq n$ particles along $\Gamma(j) \equiv \bigcup_{i=1}^j \Gamma_i$ we have
\begin{align}\label{eq:31.05}
{\cal P}\, e^{\int_{\Gamma(j)} A_{\cal S}} &= \gamma(\phi_1 = 0)\, \gamma^{-1}(\phi_j = 2\pi) \nonumber\\ 
&= \Pi_1 \ldots \Pi_j \left(\mathbbm{1} + \tfrac{1}{\kappa} \Pi_j^{-1} \ldots \Pi_1^{-1} \Upsilon_1 \Pi_1 \ldots \Pi_j + \ldots + \tfrac{1}{\kappa} \Pi_j^{-1} \Upsilon_j \Pi_j\right)\,.
\end{align}
It confirms that the composition rule is determined by the non-Abelian group multiplication (\ref{eq:21.06a}), leading to the deformed addition of both ${\bf p}_i$ and ${\bf j}_i$. When (\ref{eq:31.05}) is calculated along the whole boundary $\Gamma$, we naturally interpret $\Pi \equiv \Pi_1 \ldots \Pi_n$ as the total momentum of the system and $\Pi^{-1} \Upsilon\, \Pi \equiv \Pi_n^{-1} \Upsilon_n \Pi_n + \Pi_n^{-1} \Pi_{n-1}^{-1} \Upsilon_{n-1} \Pi_{n-1} \Pi_n + \ldots$ as the total angular momentum conjugated by $\Pi$. By construction, for a closed topology of ${\cal S}$ it has to be $\Pi = \mathbbm{1}$, $\Upsilon = 0$. The holonomy (\ref{eq:31.05}) also depends on the ordering of particles. This peculiar property \cite{Carlip:1989ey} is actually a natural feature of non-Abelian field theories in two spatial dimensions, where a system of topological defects is not invariant under a usual permutation of the pair characterized by group elements $g_i$ and $g_{i+1}$: $(g_i,g_{i+1}) \rightarrow (g_{i+1},g_i)$, but instead under a so-called braid: right-handed $(g_i,g_{i+1}) \rightarrow (g_{i+1},g_{i+1}^{-1} g_i g_{i+1})$, i.e.
\begin{align}\label{eq:31.06}
(\Pi_i,\Pi_{i+1}) &\rightarrow \left(\Pi_{i+1},\Pi_{i+1}^{-1} \Pi_i \Pi_{i+1}\right)\,, \nonumber\\ 
(\Upsilon_i,\Upsilon_{i+1}) &\rightarrow \left(\Upsilon_{i+1},{\rm Ad}(\Pi_{i+1}^{-1}) \left(\Upsilon_i - (1 - {\rm Ad}(\Pi_i))\, \Upsilon_{i+1}\right)\right)
\end{align}
or left-handed $(g_i,g_{i+1}) \rightarrow (g_i g_{i+1} g_i^{-1},g_i)$, i.e.
\begin{align}\label{eq:31.07}
(\Pi_i,\Pi_{i+1}) &\rightarrow \left(\Pi_i \Pi_{i+1} \Pi_i^{-1},\Pi_i\right)\,, \nonumber\\ 
(\Upsilon_i,\Upsilon_{i+1}) &\rightarrow \left({\rm Ad}(\Pi_i)\, \Upsilon_{i+1} + (1 - {\rm Ad}(\Pi_i \Pi_{i+1} \Pi_i^{-1}))\, \Upsilon_i,\Upsilon_i\right)\,,
\end{align}
for $i < n$. Simply speaking, a loop $\Gamma_i$ can not be pulled through $\Gamma_{i+1}$, or vice versa, but has to be deformed around the latter, which accordingly modifies the holonomy $g_i$ or $g_{i+1}$ (see \cite{Bais:1998tf} for the illustrations). Taking this into account, once a given particle ordering is chosen, the total holonomy (\ref{eq:31.05}) (with $j = n$) is defined unambiguously, since it is invariant under the braid group of $n$ elements. The above braid symmetry is obviously reflected in the properties of particle scattering and statistics at the quantum level \cite{Carlip:1989ey,Bais:1998tf,Bais:2002qy,Arzano:2014by}. 

Finally, considering the spinless case $\forall_i s_{(i)} = 0$, we may rewrite the Lagrangian (\ref{eq:31.03}) in the form that includes the mass shell constraints for individual particles, analogously to (\ref{eq:21.12}). Despite the presence of coupling between different particles, we find that it leads to the equations of motion
\begin{align}\label{eq:31.08}
\dot x_{(i)}^\alpha = \lambda_{(i)} \cos\frac{m_{(i)}}{2\kappa}\, p_{(i)}^\alpha\,, \qquad 
\dot p_{(i)}^\alpha = 0\,,
\end{align}
which are the same as for a single particle (\ref{eq:22.05}), reflecting the absence of local interactions in three-dimensional gravity. The derivation of these equations can conveniently be done in the recursive manner, starting from the $1$'st particle for variations of the Lagrangian with respect to positions and then from the $n$'th particle for variations with respect to momenta. Nevertheless, the particles experience a topological interaction between themselves via the sewing conditions (\ref{eq:31.02}) (see also \cite{Matschull:2001ty}).

\subsection{The $\kappa$-deformed Carroll case} \label{sec:3.2}
The other possibility that can be studied is to take the action (\ref{eq:30.01}) with $\Lambda > 0$, ${\bf n} = (\sqrt{\Lambda},0,0)$ and later perform the contraction of the gauge group to $\mathfrak{an}_{\bf n}(2)^* >\!\!\vartriangleleft {\rm AN}_{\bf n}(2)$, introduced in Subsection~\ref{sec:2.2}. Following the steps from the beginning of this Section and repeating calculations of the single particle case (\ref{eq:22.07}), we derive the effective Lagrangians for individual particles
\begin{align}\label{eq:32.01}
L_{(i)} = \kappa \left(\Pi_i \dot\Pi_i^{-1}\right)_\alpha \left({\bf x}_i\right)^\alpha + 
s_{(i)} \left(\bar{\mathfrak{s}}_i^{-1} \dot{\bar{\mathfrak{s}}}_i\right)_0\,,
\end{align}
with momentum $\Pi_i = \bar{\mathfrak{s}}_i e^{\frac{1}{\kappa} m_{(i)} \tilde S_0} \bar{\mathfrak{s}}_i^{-1}$ and position ${\bf x}_i = \bar{\mathfrak{s}}_i {\bf v}_i \bar{\mathfrak{s}}_i^{-1}$. Then, requiring the continuity conditions at the vertices $\gamma(\phi_{i+1} = 0) = \gamma(\phi_i = 2\pi)$ and partially fixing the gauge via $\gamma(\phi_1 = 0) = 1$, we again obtain the relations
\begin{align}\label{eq:32.02}
\mathfrak{r}_1 \bar{\mathfrak{s}}_1^{-1} &= 1\,, & 
\mathfrak{r}_2 \bar{\mathfrak{s}}_2^{-1} &= \Pi_1\,, & 
\mathfrak{r}_3 \bar{\mathfrak{s}}_3^{-1} &= \Pi_1 \Pi_2\,,\,\ldots\,, \nonumber\\ 
\bar{\mathfrak{s}}_1 {\bf v}_1 \bar{\mathfrak{s}}_1^{-1} &= {\bf x}_1\,, & 
\bar{\mathfrak{s}}_2 {\bf v}_2 \bar{\mathfrak{s}}_2^{-1} &= {\bf x}_2 + \tfrac{1}{\kappa} \Upsilon_1\,, & 
\bar{\mathfrak{s}}_3 {\bf v}_3 \bar{\mathfrak{s}}_3^{-1} &= {\bf x}_3 + \tfrac{1}{\kappa} \Upsilon_2 + \Pi_2^{-1} \tfrac{1}{\kappa} \Upsilon_1 \Pi_2\,,\,\ldots\,,
\end{align}
where the (quasi-)angular momentum $\Upsilon_i = \kappa\, ({\rm Ad}(\Pi_i^{-1}) - 1)\, {\bf x}_i + s_{(i)} \tilde J_0$ and now ${\bf x}_i \equiv \bar{\bf u}_i$. After (\ref{eq:32.02}) is applied to every $L_{(i)}$ to replace the variables ${\bf v}_i$ with $\bar{\bf u}_i$, summing over $i$ we ultimately arrive at the effective $n$-particle Lagrangian
\begin{align}\label{eq:32.03}
L_n &= \sum_{i=1}^n \left(\left(\dot{\bar{\mathfrak{s}}}_i \bar{\mathfrak{s}}_i^{-1}\right)_\alpha \left(\Upsilon_i\right)^\alpha + \left(\Pi_i \dot\Pi_i^{-1}\right)_\alpha \left(\Upsilon_{i-1} + \ldots + \Pi_{i-1}^{-1} \ldots \Pi_2^{-1} \Upsilon_1 \Pi_2 \ldots \Pi_{i-1}\right)^\alpha\right) \nonumber\\ 
&= \sum_{i=1}^n \left(\left(\Pi_i \dot\Pi_i^{-1}\right)_\alpha \left(\kappa\, {\bf x}_i + \Upsilon_{i-1} + \ldots + \Pi_{i-1}^{-1} \ldots \Pi_2^{-1} \Upsilon_1 \Pi_2 \ldots \Pi_{i-1}\right)^\alpha + s_{(i)} \left(\bar{\mathfrak{s}}_i^{-1} \dot{\bar{\mathfrak{s}}}_i\right)_0\right)\,.
\end{align}
This expression can also be arranged analogously to (\ref{eq:31.03}), as we show in the simple example
\begin{align}\label{eq:32.04}
L_3 = \sum_{i=1}^3 \left(\kappa \left(\Pi_i \dot\Pi_i^{-1}\right)_\alpha \left({\bf x}_i\right)^\alpha + s_{(i)} \left(\bar{\mathfrak{s}}_i^{-1} \dot{\bar{\mathfrak{s}}}_i\right)_0\right) + \left(\Pi_3 \dot\Pi_3^{-1}\right)_\alpha \left(\Upsilon_2\right)^\alpha + \left(\Pi_2 \Pi_3 \dot\Pi_3^{-1} \Pi_2^{-1} + \Pi_2 \dot\Pi_2^{-1}\right)_\alpha \left(\Upsilon_1\right)^\alpha\,.
\end{align}
Repeating what we did in (\ref{eq:22.13}), for vanishing spins $\forall_i s_{(i)} = 0$ we may add to (\ref{eq:32.03}) the mass shell constraint for every particle. This leads us to the equations of motion
\begin{align}\label{eq:32.05}
\dot x_{(i)}^0 = \lambda_{(i)} m_{(i)}\,, \qquad \dot x_{(i)}^a = 0\,, \qquad \dot p_{(i)}^\alpha = 0\,,
\end{align}
which are again the same as in the single particle case (\ref{eq:22.14}). In the recursive derivation of these equations we start from the $n$'th particle for variations of the Lagrangian with respect to positions and then from the $1$'st particle for variations with respect to momenta. 

Moreover, calculating the holonomy of $A_{\cal S}$ along $\Gamma(j) \equiv \bigcup_{i=1}^j \Gamma_i$, $j \leq n$ we find
\begin{align}\label{eq:32.06}
{\cal P}\, e^{\int_{\Gamma(j)} A_{\cal S}} &= \gamma(\phi_1 = 0)\, \gamma^{-1}(\phi_j = 2\pi) \nonumber\\ 
&= \left(\mathbbm{1} + \tfrac{1}{\kappa} \Pi_1 \Upsilon_1 \Pi_1^{-1} + \ldots + \tfrac{1}{\kappa} \Pi_1 \ldots \Pi_j \Upsilon_j \Pi_j^{-1} \ldots \Pi_1^{-1}\right) \Pi_1 \ldots \Pi_j\,,
\end{align}
which is determined by the group multiplication (\ref{eq:22.02a}) and allows to call $\Pi \equiv \Pi_1 \ldots \Pi_n$ the total momentum and $\Upsilon \equiv \Upsilon_n + \Pi_n^{-1} \Upsilon_{n-1} \Pi_n + \ldots$ the total (quasi-)angular momentum of all particles. For a closed ${\cal S}$ there is $\Pi = \mathbbm{1}$, $\Upsilon = 0$. The holonomy (\ref{eq:32.06}) is invariant under the braids of individual holonomies but here the transformations of angular momenta from (\ref{eq:31.06}) and (\ref{eq:31.07}) have the form
\begin{align}\label{eq:32.07}
(\Upsilon_i,\Upsilon_{i+1}) &\rightarrow \left(\Upsilon_{i+1},{\rm Ad}(\Pi_{i+1}^{-1})\, \Upsilon_i + (1 - {\rm Ad}(\Pi_{i+1}^{-1} \Pi_i^{-1} \Pi_{i+1}))\, \Upsilon_{i+1}\right)
\end{align}
and
\begin{align}\label{eq:32.08}
(\Upsilon_i,\Upsilon_{i+1}) &\rightarrow \left({\rm Ad}(\Pi_i) \left(\Upsilon_{i+1} - (1 - {\rm Ad}(\Pi_{i+1}^{-1}))\, \Upsilon_i\right),\Upsilon_i\right)\,,
\end{align}
respectively. 

We note that, as in the Poincar\'{e} case (\ref{eq:31.03}), the final Lagrangian (\ref{eq:32.03}) consists of both free and interacting terms. The only difference is that now angular momentum of a given particle is coupling to the total momentum of the following particles, instead of the preceding ones. This is associated with the fact that in the initial Lagrangians (\ref{eq:31.01}) we have a right action of the momentum sector of the gauge group on $\mathbbm{R}^{2,1}$, while for (\ref{eq:32.01}) it is a left action. The Lagrangian (\ref{eq:31.03}) can be transformed into the expression analogous to (\ref{eq:32.03}) through the following change of variables
\begin{align}\label{eq:32.09}
\Pi_i &\rightarrow \Pi_1^{-1} \dots \Pi_{i-1}^{-1} \Pi_i \ldots \Pi_1\,, \nonumber\\ 
{\bf x}_i &\rightarrow \Pi_1^{-1} \dots \Pi_{i-1}^{-1} {\bf x}_i \Pi_{i-1} \ldots \Pi_1 - \tfrac{1}{\kappa} \Pi_1^{-1} \dots \Pi_{i-1}^{-1} \Upsilon_{i-1} \Pi_{i-1} \ldots \Pi_1 - \ldots - \tfrac{1}{\kappa} \Pi_1^{-1} \Upsilon_1 \Pi_1\,, \nonumber\\ 
\bar{\mathfrak{u}}_i &\rightarrow \Pi_1^{-1} \dots \Pi_{i-1}^{-1} \bar{\mathfrak{u}}_i\,,
\end{align}
while the corresponding transformation for (\ref{eq:32.03}) is
\begin{align}\label{eq:32.10}
\Pi_i &\rightarrow \Pi_1^{-1} \dots \Pi_{i-1}^{-1} \Pi_i \ldots \Pi_1\,, \nonumber\\ 
{\bf x}_i &\rightarrow \Pi_1^{-1} \dots \Pi_{i-1}^{-1} {\bf x}_i \Pi_{i-1} \ldots \Pi_1 - \tfrac{1}{\kappa} \Pi_1^{-1} \dots \Pi_{i-2}^{-1} \Upsilon_{i-1} \Pi_{i-2} \ldots \Pi_1 - \ldots - \tfrac{1}{\kappa} \Upsilon_1\,, \nonumber\\ 
\bar{\mathfrak{s}}_i &\rightarrow \Pi_1^{-1} \dots \Pi_{i-1}^{-1} \bar{\mathfrak{s}}_i\,.
\end{align}

\section{Summary} \label{sec:4.0}
In this paper we considered the Chern-Simons theory describing gravity in three spacetime dimensions, with a system of point particles. Our approach was to apply the local factorization of gauge groups into the product of the Lorentz group and the ${\rm AN}_{\bf n}(2)$ group. This allowed us to solve the sewing condition for the Cartan connection and obtain partial results for the effective (single) particle actions with the (anti-)de Sitter group as well as with the double-product factorization of the Poincar\'{e} group. However, it remains an open question how to perform the final integration in these actions, while another complication is that the variables of particle's momentum and angular momentum belong to the subalgebras spanned by the generators $J_\alpha$ and $P_\alpha$, instead of $J_\alpha$ and $S_\alpha$. It also turns out to be similarly problematic to simplify the expressions for the holonomy of a particle in all the above cases, using the standard gauge fixing condition $\gamma(\phi = 0) = 1$ at the boundary of a punctured disc. 

On the other hand, we reviewed the known final results for particles in the theory with the Poincar\'{e} gauge group, trying to stress certain of their aspects. Furthermore, we extended the analysis of the so-called reciprocal contraction of the de Sitter group (introduced by us in an earlier paper), including the generalization to a system of multiple particles, which have not been discussed before. In this way we showed that, apart from the different type of the mass shell condition, it is completely analogous to the Poincar\'{e} case. 

Finally, let us speculate about possibilities for the related future research. An area that especially deserves more interest are applications of the Chern-Simons theory in the context of general relativity in four spacetime dimensions. In particular, the latter can also be expressed \cite{Freidel:2005qs} as a topological gauge field theory but with an extra term that breaks down the full gauge symmetry to the Lorentz symmetry, restoring local degrees of freedom. There has been a partial attempt \cite{Kowalski:2008ey} to use the Chern-Simons theory to describe a system of point particles coupled to four-dimensional gravity (with positive cosmological constant) in the limit where this gauge symmetry is preserved. A different potential research direction are (planar) gravitational waves, which effectively are three-dimensional objects, while the appropriately reduced Chern-Simons theory can be defined on a hypersurface in four dimensions \cite{Duval:1994cs}. 

One can also notice a certain similarity between three-dimensional gravity and relativistic physics in the Carrollian limit (for any number of dimensions). Namely, while in the first case there are no local interactions between particles coupled to the theory, in the second one the worldlines of particles turn out to be causally disconnected. However, the latter apparently changes when we introduce some interaction potential \cite{Bergshoeff:2014ds} and the considered analogy is probably superficial. On the other hand, as we mentioned in Subsection~\ref{sec:2.2}, spacetime in the Carrollian limit can be embedded as a null hypersurface in one dimension higher. A manifestation of this fact is that the three-dimensional Carroll group contains the symmetries of a gravitational wave in four dimensions \cite{Duval:2017cs}. The hybrid model of $\kappa$-deformed Carroll particles derived in our paper may also be worth to explore as such a link between the three and four dimensions, perhaps in the quantum context. We note here that its effective action (\ref{eq:22.13}) can be straightforwardly generalized to higher dimensions, in contrast to (\ref{eq:21.12}). 

Three-dimensional gravity naturally serves as a testing ground for different ideas associated with the quantization of general relativity. In particular, in light of the recent results that were discussed in the Introduction, the issue to be verified is whether the $\kappa$-Poincar\'{e} algebra actually describes symmetries of the effective theory of quantum gravity with particles included (as already tentatively analyzed in \cite{Cianfrani:2016ss}). Then the ${\rm AN}(2)$ group would play the role of curved momentum space, which is a characteristic ingredient in the hypothetical relative locality regime of quantum gravity \cite{Gubitosi:2013re}. However, since the above result is presumably derived in the limit of vanishing cosmological constant, it is not yet clear how it can be reconciled with the corresponding classical case, reviewed in our Subsection~\ref{sec:2.1}. The situation is additionally complicated by the peculiar case from Subsection~\ref{sec:2.2}, which is obtained instead of the expected classical particle with the ${\rm AN}(2)$ momentum space. On the other hand, the $\kappa$-deformed Carroll particles can be treated as another example of the model with curved momentum space. Furthermore, the multiparticle dynamics presented in this paper may help to improve the formulation of the principle of relative locality, as it was attempted in \cite{Kowalski:2015my}. On a separate note, let us mention that the quantum statistics satisfying the braid symmetry, called the non-Abelian anyonic statistics, can theoretically be realized in the fractional quantum Hall effect and on spin lattices, as well as be applied in the topological quantum computing \cite{Nayak:2008nn}. This allows us to establish links between these areas and gravity, see e.g.\! \cite{Pithis:2015as} (which also shows how the Chern-Simons theory can model a black hole horizon in the quantum theory) and references therein.

\section*{Acknowledgments}
The author thanks J.~Kowalski-Glikman and G.~Rosati for all comments and suggestions. This research was supported by the National Science Centre Poland, projects no. DEC-2011/02/A/ST2/00294 and 2014/13/B/ST2/04043.

\section*{Appendix} \label{sec:A.0}
For any value of the cosmological constant $\Lambda$, the (local) isometry group of three-dimensional spacetime is generated by the algebra \cite{Witten:1988dm}
\begin{align}\label{eq:A1.01}
[J_\alpha,J_\beta] = \epsilon_{\alpha\beta\gamma} J^\gamma\,, \qquad 
[J_\alpha,P_\beta] = \epsilon_{\alpha\beta\gamma} P^\gamma\,, \qquad 
[P_\alpha,P_\beta] = -\Lambda \epsilon_{\alpha\beta\gamma} J^\gamma\,,
\end{align}
where the first bracket defines the three-dimensional Lorentz subalgebra $\mathfrak{sl}(2,\mathbbm{R})$ (or equivalently $\mathfrak{su}(1,1)$). Introducing a formal parameter $\theta$, such that $\theta^2 = -\Lambda$, it is possible to make the identification of generators: $P_\alpha \equiv \theta J_\alpha$. Then each of the three isometry algebras becomes isomorphic to an extension of $\mathfrak{sl}(2,\mathbbm{R})$ over the Abelian ring $R_\Lambda$ (with a given $\Lambda$), whose elements have the form $a + \theta b \in R_\Lambda$, $a,b \in \mathbbm{R}$ \cite{Meusburger:2007gt}. For $\Lambda > 0$ 
we have $\theta = i \sqrt{\Lambda}$, $R_\Lambda \cong \mathbbm{C}$ but in other cases $\theta$ cannot be expressed via $\sqrt{|\Lambda|}$ and therefore is not a number. 

It is convenient to use the (double cover of the) Lorentz group ${\rm SL}(2,\mathbbm{R})$ (or ${\rm SU}(1,1)$) in the quaternionic representation. Namely, one can easily see that ${\rm SL}(2,\mathbbm{R})$ is isomorphic to the group of unit pseudo-quaternions $\mathbbm{H}^L_1$ (the Lorentzian version of the group of unit quaternions). Generators $e^\alpha$ of the algebra of pseudo-quaternions $\mathbbm{H}^L$ are defined by the relation
\begin{align}\label{eq:A1.02}
e_\alpha e_\beta = -\eta_{\alpha\beta} \mathbbm{1} + \epsilon_{\alpha\beta\gamma} e^\gamma\,,
\end{align}
with the identity element $\mathbbm{1}$. This leads to the following map $J_\alpha \rightarrow \frac{1}{2} e_\alpha$. Consequently, all isometry groups can be represented as the group $\mathbbm{H}^L_1 \cong {\rm SL}(2,\mathbbm{R})$ over a given ring $R_\Lambda$, whose elements can be parametrized as \cite{Meusburger:2008qr}
\begin{align}\label{eq:A1.03}
g = (k_3 + \theta q_3) \mathbbm{1} + (k^\alpha + \theta q^\alpha) J_\alpha\,,
\end{align}
where real-valued group coordinates satisfy the conditions $k_3 q_3 + \frac{1}{4} {\bf k} \cdot {\bf q} = 0$ and $k_3^2 - \Lambda q_3^2 + \frac{1}{4} ({\bf k}^2 - \Lambda {\bf q}^2) = 1$, so that $g$ is a unit pseudo-quaternion. In particular, for $\Lambda > 0$ the parametrization (\ref{eq:A1.03}) explicitly describes the ${\rm SL}(2,\mathbbm{C})$ group. For $\Lambda = 0$ the standard group structure ${\rm SL}(2,\mathbbm{R}) \vartriangleright\!\!< \mathbbm{R}^{2,1}$ can be recovered as a special (global) case of the factorization (\ref{eq:A1.08}). Finally, the situation is more subtle for $\Lambda < 0$. In order to recover the ${\rm SL}(2,\mathbbm{R}) \times {\rm SL}(2,\mathbbm{R})$ factorization one defines new generators $J^\pm_\alpha := 0_\pm J_\alpha$, where $0_\pm \equiv \frac{1}{2} (1 \pm \frac{\theta}{\sqrt{-\Lambda}})$ are the zero divisors of $R_\Lambda$, satisfying the relations $0_\pm^2 = 0_\pm$ and $0_\pm 0_\mp = 0$. Indeed, in terms of $J^\pm_\alpha$ the algebra (\ref{eq:A1.01}) (with $\Lambda < 0$) becomes
\begin{align}\label{eq:A1.04}
[J^\pm_\alpha,J^\pm_\beta] = \epsilon_{\alpha\beta\gamma} J_\pm^\gamma\,, \qquad [J^\pm_\alpha,J^\mp_\beta] = 0\,,
\end{align}
while group elements (\ref{eq:A1.03}) split into the pairs of $g_+,g_- \in {\rm SL}(2,\mathbbm{R})$, i.e.
\begin{align}\label{eq:A1.05}
g = 0_+ g_+ + 0_- g_- = 0_+ \left(u^+_3 \mathbbm{1} + u_+^\alpha J^+_\alpha\right) + 0_- \left(u^-_3 \mathbbm{1} + u_-^\alpha J^-_\alpha\right)\,,
\end{align}
where $u^\pm_3 = k_3 \pm \sqrt{-\Lambda}\, q_3$ and $u_\pm^\alpha = k^\alpha \pm \sqrt{-\Lambda}\, q^\alpha$. 

On the other hand, introducing the generators
\begin{align}\label{eq:A1.06}
S_\alpha := P_\alpha + \epsilon_{\alpha\beta\gamma} n^\beta J^\gamma\,, \quad 
{\bf n}^2 = \Lambda\,,
\end{align}
where ${\bf n}$ is some vector from $\mathbbm{R}^{2,1}$, we can rewrite the algebra (\ref{eq:A1.01}) as
\begin{align}\label{eq:A1.07}
[J_\alpha,J_\beta] = \epsilon_{\alpha\beta\gamma} J^\gamma\,, \qquad 
[J_\alpha,S_\beta] = \epsilon_{\alpha\beta\gamma} S^\gamma + 
n_\beta J_\alpha - \eta_{\alpha\beta} n^\gamma J_\gamma\,, \qquad 
[S_\alpha,S_\beta] = n_\alpha S_\beta - n_\beta S_\alpha\,.
\end{align}
The third bracket defines the so-called $\mathfrak{an}_{\bf n}(2)$ algebra, which can be seen as a deformed $\mathbbm{R}^{2,1}$ algebra, with the deformation vector ${\bf n}$. In the case ${\bf n} = 0$ (when $\Lambda = 0$), the algebra actually becomes $\mathbbm{R}^{2,1}$. Furthermore, it has been shown \cite{Meusburger:2008qr} that there exists the corresponding factorization of isometry group elements into
\begin{align}\label{eq:A1.08}
g = \mathfrak{u}\, \mathfrak{s} = (u_3 \mathbbm{1} + u^\alpha J_\alpha) (s_3 \mathbbm{1} + s^\beta S_\beta)\,,
\end{align}
where $\mathfrak{u} \in {\rm SL}(2,\mathbbm{R})$, $\mathfrak{s} \in {\rm AN}_{\bf n}(2)$ (or $\mathfrak{s} \in \mathbbm{R}^{2,1}$), if the condition $s_3 + \tfrac{1}{2} {\bf n} \cdot {\bf s} > 0$ is satisfied. $u_3$ and $s_3$ are given by $u_3 = \sqrt{1 - \frac{1}{4} {\bf u}^2}$, $s_3 = \sqrt{1 + \frac{1}{4} ({\bf n} \cdot {\bf s})^2}$. The relation with the global group parametrization (\ref{eq:A1.03}) is presented in \cite{Meusburger:2008qr}, although not in terms of coordinates. Here we calculate their explicit form (the same for any $\Lambda$)
\begin{align}\label{eq:A1.09}
u_3 &= \frac{1}{N_L} (k_3 + \tfrac{1}{2} {\bf n} \cdot {\bf q})\,,& u^\alpha &= \frac{1}{N_L} \left(k^\alpha - 2q_3 n^\alpha + \epsilon^\alpha_{\ \beta\gamma} n^\beta q^\gamma\right)\,, \nonumber\\
s_3 &= \frac{1}{2N_L} (N_L^2 + 1)\,,& 
s^\alpha &= \frac{1}{N_L} \left(k_3 q^\alpha - q_3 k^\alpha - \tfrac{1}{2} \epsilon^\alpha_{\ \beta\gamma} k^\beta q^\gamma + 2 (q_3^2 + \tfrac{1}{4} {\bf q}^2) n^\alpha\right)\,,
\end{align}
where the normalizing constant 
\begin{align}\label{eq:A1.10}
N_L^2 \equiv k_3^2 + \Lambda q_3^2 + \tfrac{1}{4} ({\bf k}^2 + \Lambda {\bf q}^2) + k_3 {\bf n} \cdot {\bf q} - q_3 {\bf n} \cdot {\bf k} - \tfrac{1}{2} \epsilon_{\alpha\beta\gamma} n^\alpha k^\beta q^\gamma\,,
\end{align}
which is equivalent to $N_L = s_3 + \frac{1}{2} {\bf n} \cdot {\bf s}$. 

On the other hand, a group element can be factorized in the reverse order into
\begin{align}\label{eq:A1.11}
g = \mathfrak{r}\, \mathfrak{v} = (r_3 \mathbbm{1} + r^\alpha S_\alpha) (v_3 \mathbbm{1} + v^\beta J_\beta)\,,
\end{align}
where $\mathfrak{r} \in {\rm AN}_{\bf n}(2)$, $\mathfrak{v} \in {\rm SL}(2,\mathbbm{R})$ (or $\mathfrak{r} \in \mathbbm{R}^{2,1}$), under the condition $r_3 - \tfrac{1}{2} {\bf n} \cdot {\bf r} > 0$. $v_3$ and $r_3$ are again given by $v_3 \equiv \sqrt{1 - \frac{1}{4} {\bf v}^2}$, $r_3 \equiv \sqrt{1 + \frac{1}{4} ({\bf n} \cdot {\bf r})^2}$. The expressions for coordinates of (\ref{eq:A1.11}) are similar to the previous ones (\ref{eq:A1.09}). Furthermore, when both factorizations exist, it is naturally possible to make a transformation from the first to the second one or vice versa \cite{Meusburger:2008qr}. In the former case we calculate here the following explicit formulae
\begin{align}\label{eq:A1.12}
\mathfrak{v} = \frac{1}{N_R} (\nu_3 \mathbbm{1} + \nu^\alpha J_\alpha)\,, \quad 
\nu_3 &= u_3 \left(s_3 - \tfrac{1}{2} {\bf n} \cdot {\bf s}\right) + \tfrac{1}{2} \epsilon_{\alpha\beta\gamma} u^\alpha n^\beta s^\gamma\,, \nonumber\\ 
\nu^\alpha &= \left(s_3 - \tfrac{1}{2} {\bf n} \cdot {\bf s}\right) u^\alpha + {\bf n} \cdot {\bf s}\, u^\alpha - {\bf u} \cdot {\bf s}\, n^\alpha 
\end{align}
and
\begin{align}\label{eq:A1.13}
\mathfrak{r} = \frac{1}{N_R} (\varrho_3 \mathbbm{1} + \varrho^\alpha S_\alpha)\,, \quad 
\varrho_3 &= 1 + \tfrac{1}{8} {\bf s}^2 \left({\bf n}^2 {\bf u}^2 - ({\bf n} \cdot {\bf u})^2\right) \nonumber\\ 
&+ \tfrac{1}{2} \left(s_3 - \tfrac{1}{2} {\bf n} \cdot {\bf s} \right) \left(-{\bf n} \cdot {\bf s} + \tfrac{1}{2} \left({\bf n} \cdot {\bf s}\, {\bf u}^2 - {\bf n} \cdot {\bf u}\, {\bf u} \cdot {\bf s}\right) + u_3 \epsilon_{\alpha\beta\gamma} u^\alpha n^\beta s^\gamma\right)\,, \nonumber\\ 
\varrho^\alpha &= \tfrac{1}{2} {\bf s}^2 \left(\tfrac{1}{2} \left({\bf n} \cdot {\bf u}\, u^\alpha - {\bf u}^2 n^\alpha\right) - u_3 \epsilon^\alpha_{\ \beta\gamma} n^\beta u^\gamma\right) \nonumber\\ 
&+ \left(s_3 - \tfrac{1}{2} {\bf n} \cdot {\bf s} \right) \left(s^\alpha + \tfrac{1}{2} \left({\bf u} \cdot {\bf s}\, u^\alpha - {\bf u}^2 s^\alpha\right) + u_3 \epsilon^\alpha_{\ \beta\gamma} u^\beta s^\gamma\right)\,.
\end{align}
The normalizing constant $N_R$ can be written as $N_R = r_3 - \tfrac{1}{2} {\bf n} \cdot {\bf r}$, with the condition $r_3 - \tfrac{1}{2} {\bf n} \cdot {\bf r} > 0$ or explicitly
\begin{align}\label{eq:A1.14}
N_R^2 \equiv 1 + \tfrac{1}{4} {\bf s}^2 \left({\bf n}^2 {\bf u}^2 - ({\bf n} \cdot {\bf u})^2\right) + \left(s_3 - \tfrac{1}{2} {\bf n} \cdot {\bf s} \right) \left(-{\bf n} \cdot {\bf s} + \tfrac{1}{2} \left({\bf n} \cdot {\bf s}\, {\bf u}^2 - {\bf n} \cdot {\bf u}\, {\bf u} \cdot {\bf s}\right) + u_3 \epsilon_{\alpha\beta\gamma} u^\alpha n^\beta s^\gamma\right)\,,
\end{align}
and by construction it satisfies the relations $N_R^2 = \varrho_3^2 - \tfrac{1}{4} (n^\alpha \varrho_\alpha)^2$, $\varrho_3 > \tfrac{1}{2}$ as well as $N_R^2 = \nu_3^2 + \tfrac{1}{4} \nu^\alpha \nu_\alpha$. For clarity let us also note that $\varrho_3 + \tfrac{1}{2} n^\alpha \varrho_\alpha = 1$.

\end{document}